\documentclass[reqno,12pt]{amsart}
\usepackage{fullpage}
\usepackage{amsfonts}
\usepackage{amssymb}
\usepackage{amsmath}

\numberwithin{equation}{section}
\begin{document}

\def\Rnum{\mathbb{R}}

\def\Rop{{\mathcal R}}
\def\Jop{{\mathcal J}}
\def\Hop{{\mathcal H}}
\def\Eop{{\mathcal E}}
\def\Dop{{\mathcal D}}
\def\X{{\mathbf X}}
\def\pr{{\rm pr}}
\def\lieder{{\mathfrak L}}
\def\w{{\boldsymbol\omega}}

\def\A{A}
\def\B{B}
\def\C{C}
\def\F{F}
\def\G{G}
\def\I{I}
\def\J{J}
\def\K{K}
\def\M{M}
\def\N{N}
\def\P{P}
\def\R{R}
\def\S{S}

\def\p{\partial}
\def\const{{\rm const.}}

\def\parder#1#2{\frac{\partial{#1}}{\partial{#2}}}
\def\parderop#1{\partial/\partial{#1}}
\def\smallint{{\textstyle\int}}

\newtheorem{prop}{Proposition}
\newtheorem{thm}{Theorem}
\newtheorem{cor}{Corollary}
\newtheorem{lem}{Lemma}

\def\propref#1{Proposition~\ref{#1}}
\def\proprefs#1#2{Propositions~\ref{#1} and~\ref{#2}}
\def\thmref#1{Theorem~\ref{#1}}
\def\thmrefs#1#2{Theorems~\ref{#1} and~\ref{#2}}
\def\lemref#1{Lemma~\ref{#1}}
\def\lemrefs#1#2{Lemmas~\ref{#1} and~\ref{#2}}
\def\Ref#1{Ref.\cite{#1}}
\def\Refs#1{Refs.\cite{#1}}
\def\secref#1{Sec.~\ref{#1}}

\def\ie/{i.e.}
\def\eg/{e.g.}
\def\etc/{etc.}

\allowdisplaybreaks[4]

\title{Generalized negative flows in hierarchies of integrable evolution equations}

\author{
Stephen C. Anco$^1$, 
Shahid Mohammad$^{1,2}$,
Thomas Wolf$^1$,
Chunrong Zhu$^{1,3}$
\\
\lowercase{\scshape{
${}^1$
Department of Mathematics and Statistics, 
Brock University\\
St. Catharines, ON L2S3A1, Canada
}}\\
\lowercase{\scshape{
${}^2$
Department of Mathematics, 
Central Michigan University\\
Mount Pleasant, MI 48859, USA
}}\\
\lowercase{\scshape{
${}^3$
College of Mathematics and Computer Science,
Anhui Normal University\\
Wuhu, Anhui 241000, China
}}
}

\begin{abstract}
A one-parameter generalization of the hierarchy of negative flows is introduced 
for integrable hierarchies of evolution equations,
which yields a wider (new) class of non-evolutionary integrable nonlinear wave equations. 
As main results, 
several integrability properties of these generalized negative flow equation 
are established, 
including their symmetry structure, conservation laws, and bi-Hamiltonian formulation. 
(The results also apply to the hierarchy of ordinary negative flows). 
The first generalized negative flow equation is worked out explicitly for 
each of the following integrable equations: 
Burgers, Korteweg-de Vries, modified Korteweg-de Vries, Sawada-Kotera, Kaup-Kupershmidt, Kupershmidt. 
\end{abstract}

\maketitle

\section{Introduction}

The study of integrability of nonlinear partial differential equations (PDEs) 
has been an active area of research for several decades. 
Presently, the most direct and widest definition of integrability is that a PDE system 
should possess a hierarchy of symmetries \cite{Fok,IbrSha}. 
Other aspects of integrability possessed by some PDE systems 
are a hierarchy of conservation laws \cite{IbrSha}, 
a bi-Hamiltonian structure \cite{Mag,Olv}, 
and a Lax pair \cite{Lax} or an inverse scattering transform \cite{ZakSha}. 

For integrable scalar evolution equations in 1+1 dimensions, 
both the equation and its symmetries can be typically expressed as 
a hierarchy of flows \cite{Olv77}
\begin{equation}\label{integrhierarchy}
u_t = \Rop^n(u_x),
\quad
n=1,2,\ldots
\end{equation}
given in terms of 
a recursion operator $\Rop$, 
where $n=1$ yields the root integrable equation $u_t=\Rop(u_x)$, 
and each $n\geq 2$ yields a higher-order integrable equation 
corresponding to a symmetry in the hierarchy $\X^{(n)}= \Rop^n(u_x)\p_u$, 
while $n=0$ yields the root symmetry consisting of $x$-translations $\X^{(0)}= u_x\p_u$
in evolutionary form.

A complete classification is known 
\cite{symmintegr1,symmintegr2,symmintegr3,symmintegr4,SanWan98} 
for all hierarchies of integrable scalar evolution equations \eqref{integrhierarchy}
with the semilinear polynomial form 
$u_t=u_{mx} + f(u,u_x,\ldots,u_{(m-1)x})$, $m>1$, 
where $f$ is a polynomial function with positive scaling weight $k>0$ 
under a group of scaling transformations 
$x\rightarrow\lambda x$, $t\rightarrow\lambda^m t$, $u\rightarrow\lambda^{-q} u$
(with group parameter $\lambda\neq 0$). 
These hierarchies consist of:\\
$\bullet$ 
$m=2$ Burgers ($q=1$); \\
$\bullet$ 
$m=3$ Korteweg-de Vries (KdV) ($q=2$), 
modified Korteweg-de Vries (mKdV) ($q=1$);\\
$\bullet$
$m=5$ Sawada-Kotera (SK) ($q=2$), 
Kaup-Kupershmidt (KK) ($q=2$), 
Kupershmidt (K) ($q=1$);\\
$\bullet$
potential versions of those. \\
In particular, 
every integrable semilinear positive-weight polynomial scalar evolution equation 
except the Ibragimov-Shabat (IS) equation 
belongs to these six hierarchies. 
(Interestingly, there is a differential substitution that transforms the IS equation into 
a semilinear rational evolution equation 
which is the root equation in an integrable hierarchy.)

In contrast \cite{SanWan00}, 
no integrability classification yet exists for non-polynomial scalar evolution equations
as well as for non-evolution scalar equations. 
However, there is a class of such equations associated to any hierarchy of 
integrable scalar evolution equations by considering flows defined by the kernel of 
the underlying recursion operator. 
These flows take the form 
\begin{equation}\label{negflowhierarchy}
\Rop^n(u_t)=0,
\quad
n=1,2,\ldots
\end{equation}
which are commonly called ``negative flows'' \cite{Ver}, 
where $\Rop$ is the recursion operator of the root integrable equation $u_t=\Rop(u_x)$. 
Each negative flow can be expressed, after some differential manipulations,
as an equation of the general form 
$u_{t\,mx}=f(u,u_x,\ldots,u_{(m-1)x},u_t,u_{tx},\ldots,u_{t\,(m-1)x})$, $m>1$. 
A well-known example is the sine-Gordon equation, 
which arises as a special case of the first negative flow in the mKdV hierarchy
\cite{ZakManNovPit}. 
Another familiar example is the Camassa-Holm equation,
which arises from the first negative flow in the KdV hierarchy 
by a hodograph transformation \cite{Fuc96}.

In the present paper, 
we introduce and study a more general class of negative flows given by 
\begin{equation}\label{generalnegflowhierarchy}
\Rop^n(u_t)=\kappa_n u_t,
\quad
n=1,2,\ldots
\end{equation}
with $\kappa_n$ being a (constant) parameter. 
When $\kappa_n=0$, this class reduces the ordinary negative flows \eqref{negflowhierarchy}
associated to the root integrable equation $u_t=\Rop(u_x)$. 

We will establish the following general results in \secref{generalresults}. 
First, 
if the recursion operator $\Rop$ is hereditary \cite{Fuc96},
then each generalized negative flow equation shares 
the hierarchy of symmetries of the root integrable equation
and hence is integrable itself. 
Second, 
if the root integrable equation possesses a bi-Hamiltonian structure, 
corresponding to a factorization of the recursion operator $\Rop$, 
then each generalized negative flow equation also has a bi-Hamiltonian structure. 
Third, 
the resulting hierarchy of conservation laws arising from 
a bi-Hamiltonian structure for the root equation 
determines a related conservation law hierarchy for each generalized negative flow equation. 
Fourth, 
this structure also gives rise to spatial first integrals that can be used to integrate 
each generalized negative flow equation to obtain an equivalent flow equation 
having a lower differential order. 
All of these main results hold equally well for the ordinary negative flows \eqref{negflowhierarchy}. 

In \secref{scalarevoleqns}, 
we will work out the explicit form of the first generalized negative flow 
for each of the integrable semilinear positive-weight polynomial scalar evolution equation hierarchies:
Burgers, KdV, mKdV, SK, KK, K. 
Their respective hierarchies of symmetries, 
bi-Hamiltonian structures,  
and conservation laws are also derived. 

Finally, in \secref{remarks},
we will make some concluding remarks. 

Other work on negative flows from different points of view can be found in 
\Ref{AblKauNewSeg,CalDeg,Qia,QiaCaoStr,GomFraMelZim,FraGomZim}.

\section{General results on negative flows}
\label{generalresults}

We start with some preliminaries that will be needed in stating and proving the main results. 
The appropriate setting is variational calculus on jet spaces (see \Ref{Olv}). 

Let $u(t,x)$ be a scalar function of two independent variables $t,x$. 
The jet space associated to the set of these functions is the coordinate space defined by 
the variables $t,x,u$ and the partial derivatives of $u$ to all orders, 
$J^\infty=(t,x,u,u_t,u_x,\ldots)$. 
A differential function on $J^\infty$ is a smooth function 
$f(t,x,u,u_t,u_x,\ldots)$ of finitely many variables. 
A vector field in evolutionary form on $J^\infty$ consists of an operator $\X_\eta= \eta\p_u$ 
where $\eta$ is some differential function. 
The action of a vector field on differential functions is given by the Frechet derivative 
$\pr\X_\eta(f) = \eta\p_{u}f +(D_t\eta) \p_{u_t}f +(D_x\eta)\p_{u_x}f +\cdots = \delta_\eta f$
where $D$ denotes a total derivative and $\pr$ denotes prolongation. 
This action can be viewed naturally as the infinitesimal change in the function $f$ 
under the flow $u_t=\X_\eta(u)=\eta$ defined by the vector field $\X_\eta=\eta\p_u$. 
The Frechet derivative operator applied to a differential function is defined as 
$f{}^\prime= \p_{u}f +(\p_{u_t}f)D_t +(\p_{u_x}f)D_x +\cdots$,
whereby $f{}^\prime(\eta)= \delta_\eta f$. 
This operator is related to the Euler-Lagrange operator 
$E_u = \p_{u} -D_t\p_{u_t} -D_x\p_{u_x} +D_t^2\p_{u_{tt}} +D_tD_x\p_{u_{tx}} + D_x^2\p_{u_{xx}}+ \cdots$ 
by the variational identity 
\begin{equation}\label{varid}
f{}^\prime(\eta) - \eta E_u(f) =D_t\Phi^t(f,\eta) +D_x\Phi^x(f,\eta)
\end{equation}
where 
\begin{equation}
\begin{aligned}
& \Phi^t(f,\eta) = \eta\p_{u_t}f  +(D_t\eta)\p_{u_{tt}}f -\eta D_t\p_{u_{tt}}f+\tfrac{1}{2}(D_x\eta)\p_{u_{tx}}f  -\tfrac{1}{2}\eta D_x\p_{u_{tx}}f  +\cdots
\\
& \Phi^x(f,\eta) = \eta\p_{u_x}f  +(D_x\eta)\p_{u_{xx}}f -\eta D_x\p_{u_{xx}}f  +\tfrac{1}{2}(D_t\eta)\p_{u_{tx}}f -\tfrac{1}{2} \eta D_t\p_{u_{tx}}f +\cdots . 
\end{aligned}
\end{equation}

A linear pseudo-differential operator has the form 
\begin{equation}\label{op}
\Rop= \sum_{k=-\infty}^{q}R_kD_x^k
\end{equation}
where $q$ is a non-negative integer, 
and each $R_k$ is a differential function. 
For $k<0$, $D_x^k$ denotes $(D_x^{-1})^{|k|}$ 
where the inverse total derivative $D_x^{-1}$ is defined by the properties 
$D_x^{-1}D_x=1$ and $[D_x^{-1},D_x]=0$ 
when these operators act on an arbitrary differential function. 
The adjoint $\Rop^*$ of a linear pseudo-differential operator is defined 
through the integration by parts identity
\begin{equation}\label{adjointid}
g\Rop f - f\Rop^* g = D_x\Phi_\Rop^x(g,f)
\end{equation}
where 
\begin{equation}
\Phi^x_\Rop(g,f) = \sum_{k=0}^{q}(D_x^kf) \sum_{l=1}^{q-k} (-D_x)^{l-1}(R_{k+l}g)
+ \sum_{k=-\infty}^{-1}(D_x^kf) \sum_{l=-\infty}^{0} (-D_x)^{l-1}(R_{k+l}g) .
\end{equation}
Any linear pseudo-differential operator has a formal inverse 
which is a linear pseudo-differential operator $\Rop^{-1}$ defined by the property 
$\Rop\Rop^{-1}f=\Rop^{-1}\Rop f=f$ where $f$ is an arbitrary differential function. 
Note that a given operator $\Rop$ can have a kernel $f$ whereby 
$\Rop f=0$ holds for some particular differential function $f$ which is possibly nonlocal 
(namely, $f$ is some function of $t,x$, derivatives of $u$, and nonlocal variables given by 
$D_x^ku$ and $D_x^ku_t$ with $k<0$). 

The Frechet derivative of a linear pseudo-differential operator is defined 
coefficient-wise by 
$\delta_\eta\Rop = \sum_{k=-\infty}^{q}(\delta_\eta R_k)D_x^k$. 
A vector field $\X_\eta$ acts on a linear pseudo-differential operator by the Lie derivative 
\begin{equation}
\lieder_{\X_\eta}\Rop = \pr\X_\eta(\Rop) +[\Rop,\eta{}^\prime]
\end{equation}
where the bracket denotes the commutator of operators. 
This action arises naturally as the infinitesimal change in the operator $\Rop$ 
under the flow $u_t=\X_\eta(u)=\eta$ defined by the vector field $\X_\eta=\eta\p_u$. 

Note that for $\eta=-u_t$ the vector field $\X_\eta=-u_t\p_u$ 
generates time translations, 
and consequently $\lieder_{\X_\eta}\Rop = -\delta_{u_t}\Rop +[D_t,\Rop]=\Rop_t$ 
acts as the partial $t$-derivative on the coefficients of $\Rop$. 
The operator $\Rop$ is thereby invariant under time translations iff 
\begin{equation}\label{timetransinv}
\Rop_t=0 . 
\end{equation}

For the results hereafter, 
the main condition that will be needed on linear pseudo-differential operators $\Rop$
is the {\em hereditary property} \cite{Fuc96} 
\begin{equation}\label{hereditaryprop}
\lieder_{\X_{\Rop\eta}}\Rop = \Rop(\lieder_{\X_{\eta}}\Rop)
\end{equation}
holding for all differential functions $\eta$. 
An equivalent formulation is given by 
\begin{equation}\label{hereditaryop}
\delta_{\Rop\eta}\Rop +[\Rop,(\Rop\eta){}^\prime]= \Rop(\delta_{\eta}\Rop+[\Rop,\eta{}^\prime]) . 
\end{equation}

\subsection{Symmetry properties and integrability of negative flows}

Generalized negative flow equations have the form 
\begin{equation}\label{generalflow}
F = \Rop^n(u_t) - \kappa_n u_t =0,
\quad
\kappa_n=\const
\end{equation}
in terms of a linear pseudo-differential operator $\Rop$, 
where $n$ is a positive integer and $\kappa_n$ is a constant. 
In particular, equation \eqref{generalflow} will be called the {\em generalized $-n$ flow}.

A vector field $\X=\eta\p_u$ is an infinitesimal symmetry of this equation \eqref{generalflow} 
iff it leaves the solution space of the equation invariant, 
$(\pr\X F)|_{F=0} =0$. 
This invariance condition can be formulated equivalently as 
\begin{equation}\label{symmdeteqn}
\begin{aligned}
0=(\delta_{\eta} F)|_{F=0}
& = \left( \delta_{\eta} (\Rop^n u_t) - \kappa_n \delta_{\eta} u_t \right)|_{F=0}
\\
&= \Big( \sum_{k=1}^{n}\Rop^{k-1}((\delta_{\eta}\Rop)\Rop^{n-k}u_t)
+\Rop^n(D_t\eta) - \kappa_n D_t\eta \Big)\Big|_{F=0}
\end{aligned}
\end{equation}
which constitutes a determining equation on the differential function $\eta$. 

To begin, we first establish conditions for $\Rop$ to be a symmetry recursion operator. 

\begin{lem}\label{recursop}
If the operator $\Rop$ is both hereditary \eqref{hereditaryprop} 
and time-translation invariant \eqref{timetransinv}, 
then $\Rop$ is a symmetry recursion operator for the flow equation \eqref{generalflow}.
\end{lem}

The proof is straightforward. 
It is necessary and sufficient to show that if 
$\X_{\eta}=\eta\p_u$ is an infinitesimal symmetry 
then so is $\X_{\Rop\eta}=(\Rop\eta)\p_u$. 
Consider 
\begin{equation}\label{varRetaF}
\delta_{\Rop\eta} F 
= \delta_{\Rop\eta} (\Rop^n u_t) - \kappa_n \delta_{\Rop\eta} u_t
= \sum_{k=1}^{n}\Rop^{k-1}((\delta_{\Rop\eta} \Rop)\Rop^{n-k}u_t) +\Rop^n(D_t(\Rop\eta)) - \kappa_n D_t(\Rop\eta) . 
\end{equation}
The first term in this expression \eqref{varRetaF} can be expanded by using 
the hereditary property \eqref{hereditaryop} to get 
\begin{equation}\label{varRetaF-term1}
\begin{aligned}
\Rop^{k-1}((\delta_{\Rop\eta} \Rop)\Rop^{n-k}u_t) 
& = 
\Rop^{k}((\delta_\eta\Rop)\Rop^{n-k}u_t) 
+\Rop^{k+1}(\delta_{\Rop^{n-k}u_t}\eta) -\Rop^{k}(\delta_{\Rop^{n-k+1}u_t}\eta)
\\&\qquad
+\Rop^{k-1}(\delta_{\Rop^{n-k+1}u_t}(\Rop\eta)) -\Rop^{k}(\delta_{\Rop^{n-k}u_t}(\Rop\eta)) .
\end{aligned}
\end{equation}
In this expanded expression \eqref{varRetaF-term1},
the first term simplifies through the symmetry equation \eqref{symmdeteqn}
\begin{equation}
\Big( \sum_{k=1}^{n} \Rop^{k}((\delta_\eta\Rop)\Rop^{n-k}u_t) \Big)\Big|_{F=0}
= -\Rop^{n+1}(D_t\eta) +\kappa_n \Rop(D_t\eta)
\end{equation}
while the remaining terms 
yield telescoping sums 
\begin{equation}\label{sum1}
\sum_{k=1}^{n} \Big( \Rop^{k+1}(\delta_{\Rop^{n-k}u_t}\eta) -\Rop^{k}(\delta_{\Rop^{n-k+1}u_t}\eta) \Big)
= \Rop^{n+1}(\delta_{u_t}\eta) -\Rop(\delta_{\Rop^{n}u_t}\eta) 
\end{equation}
and
\begin{equation}\label{sum2}
\sum_{k=1}^{n} \Big( \Rop^{k-1}(\delta_{\Rop^{n-k+1}u_t}(\Rop\eta)) -\Rop^{k}(\delta_{\Rop^{n-k}u_t}(\Rop\eta)) \Big)
= \delta_{\Rop^{n}u_t}(\Rop\eta) -\Rop^{n}(\delta_{u_t}(\Rop\eta)) .
\end{equation}
After the flow equation \eqref{generalflow} is substituted into the terms \eqref{sum1} and \eqref{sum2}, 
the expanded expression \eqref{varRetaF-term1} yields 
\begin{equation}\label{varRetaF-term1-simplified}
\begin{aligned}
\Big( \sum_{k=1}^{n} \Rop^{k-1}((\delta_{\Rop\eta} \Rop)\Rop^{n-k}u_t) \Big)\Big|_{F=0}
& = 
\Rop^{n+1}(\delta_{u_t}\eta)  -\Rop^{n+1}(D_t\eta) 
+\kappa_n \Rop(D_t\eta) -\kappa_n\Rop(\delta_{u_t}\eta) 
\\&\qquad
+\kappa_n \delta_{u_t}(\Rop\eta) -\Rop^{n}(\delta_{u_t}(\Rop\eta)) . 
\end{aligned}
\end{equation}
Then this expression \eqref{varRetaF-term1-simplified} 
combines with the second and third terms in expression \eqref{varRetaF} to give 
\begin{equation}\label{varRetaF-simplified}
\begin{aligned}
&\Big( \sum_{k=1}^{n}\Rop^{k-1}((\delta_{\Rop\eta} \Rop)\Rop^{n-k}u_t) +\Rop^{n}(D_t(\Rop\eta)) - \kappa_n D_t(\Rop\eta) \Big)\Big|_{F=0}
\\&
= \Rop^{n}(D_t(\Rop\eta) -\delta_{u_t}(\Rop\eta))
+ \Rop^{n+1}(\delta_{u_t}\eta -D_t\eta) 
\\&\qquad
+\kappa_n \Rop(D_t\eta -\delta_{u_t}\eta) 
+\kappa_n(\delta_{u_t}(\Rop\eta) - D_t(\Rop\eta)) . 
\end{aligned}
\end{equation}
Each term now simplifies due to the relation $D_tf - \delta_{u_t}f = f_t$ 
which holds for any differential function $f$. 
This yields
\begin{equation}
\Rop^{n}(D_t(\Rop\eta) -\delta_{u_t}(\Rop\eta))
+ \Rop^{n+1}(\delta_{u_t}\eta -D_t\eta) 
= \Rop^{n}(\Rop_t\eta)
\end{equation}
and
\begin{equation}
\kappa_n \Rop(D_t\eta -\delta_{u_t}\eta) +\kappa_n(\delta_{u_t}(\Rop\eta) - D_t(\Rop\eta))
= -\kappa_n \Rop_t\eta . 
\end{equation}
Hence, with these simplifications, the expression \eqref{varRetaF} becomes
\begin{equation}
\Big( \sum_{k=1}^{n}\Rop^{k-1}((\delta_{\Rop\eta} \Rop)\Rop^{n-k}u_t) +\Rop^{n}(D_t(\Rop\eta)) - \kappa_n D_t(\Rop\eta) \Big)\Big|_{F=0}
= \Rop^{n}(\Rop_t\eta)-\kappa_n \Rop_t\eta
\end{equation}
which shows that $(\delta_{\Rop\eta} F)|_{F=0}= \Rop^n(\Rop_t\eta) - \kappa_n \Rop_t\eta$
will vanish whenever $\Rop_t=0$. 
This completes the proof. 

Next we establish an invariance condition on $\Rop$ 
for the generalized negative flow equation \eqref{generalflow} to possess 
space translations as symmetries. 
Recall, space translations are generated by the vector field $\X_\eta=-u_x\p_u$ with $\eta=-u_x$. 

\begin{lem}\label{xtranssymm}
An operator $\Rop$ is invariant under space translations iff 
\begin{equation}\label{xtransinvop}
0=\lieder_{\X_\eta}\Rop = -\delta_{u_x}\Rop +[D_x,\Rop]=\Rop_x
\end{equation}
For any space-translation invariant operator $\Rop$, 
the corresponding flow equation \eqref{generalflow} 
possesses space translations $\X_\eta=-u_x\p_u$ as symmetries. 
\end{lem}

The proof consists of showing that  $\eta=-u_x$ 
satisfies the symmetry equation \eqref{symmdeteqn},
without use of the hereditary property \eqref{hereditaryop}. 
Consider 
\begin{equation}\label{varuxF}
\delta_{\eta} F 
= -\delta_{u_x} (\Rop^n u_t) + \kappa_n \delta_{u_x} u_t
= -\sum_{k=1}^{n}\Rop^{k-1}((\delta_{u_x} \Rop)\Rop^{n-k}u_t) -\Rop^n(D_t(u_x)) + \kappa_n D_t(u_x) . 
\end{equation}
The first term in this expression \eqref{varuxF} can be expanded by using 
the invariance property \eqref{xtransinvop} to get 
\begin{equation}
-\Rop^{k-1}((\delta_{u_x} \Rop)\Rop^{n-k}u_t) 
= \Rop^{k}(D_x(\Rop^{n-k}u_t)) - \Rop^{k-1}(D_x(\Rop^{n-k+1}u_t)) 
\end{equation}
which yields a telescoping sum 
\begin{equation}\label{varuxF-term1}
\sum_{k=1}^{n} \Big( \Rop^{k}(D_x(\Rop^{n-k}u_t)) - \Rop^{k-1}(D_x(\Rop^{n-k+1}u_t)) \Big)
= \Rop^{n}(D_xu_t) - D_x(\Rop^{n}u_t) . 
\end{equation}
Substitution of the flow equation \eqref{generalflow} into this expression \eqref{varuxF-term1}
then shows that the first term in expression \eqref{varuxF} becomes 
\begin{equation}\label{varuxF-term1-simplified} 
\Big(-\sum_{k=1}^{n} \Rop^{k-1}((\delta_{u_x} \Rop)\Rop^{n-k}u_t) \Big)\Big|_{F=0}
= \Rop^{n}u_{tx} -\kappa_n u_{tx} . 
\end{equation}
Now this expression \eqref{varuxF-term1-simplified} can be combined 
with the second and third terms in expression \eqref{varuxF}, 
yielding $(\delta_{\eta} F)|_{F=0}= 0$.
This completes the proof. 

We can now state the main integrability result by combining \lemrefs{recursop}{xtranssymm} 
along with the following standard property of hereditary operators \cite{Olv}:
if $\Rop$ is a hereditary operator, then it is a recursion operator 
for each evolution equation in the hierarchy of flows \eqref{integrhierarchy}. 

\begin{thm}\label{symmsnegflow}
Let $\Rop$ be a hereditary operator that is invariant under both time and space translations. 
Then the flow equation \eqref{generalflow} possesses an infinite hierarchy of symmetries
\begin{equation}\label{symmhierarchy}
\X^{(k)} = -\Rop^k(u_x)\p_u, 
\quad
k=0,1,2,\ldots
\end{equation}
starting with the space translation symmetry $\X^{(0)} = -u_x\p_u$.
The next symmetry $\X^{(1)} = -\Rop(u_x)\p_u$ 
can be viewed as defining a root flow equation $u_t=\Rop(u_x)$
for which the equation \eqref{generalflow} represents a generalized negative flow
for every $n=1,2,\ldots$. 
In particular, 
all of the generalized negative flows, 
the root flow, 
and all of the higher-order flows defined by $u_t=\Rop^k(u_x)$, $k=2,3,\ldots$, 
comprise a hierarchy of flows with the integrability property that 
they share all of the symmetries \eqref{symmhierarchy}. 
\end{thm}

\subsection{Bi-Hamiltonian structure of negative flows}

A Hamiltonian operator $\Dop$ is a linear pseudo-differential operator that 
is skew and obeys a Jacobi relation. 
These two properties have an equivalent formulation in terms of 
an associated Poisson bracket defined by 
\begin{equation}\label{poissonbrack}
\{\mathfrak{F_1},\mathfrak{F_2}\}_\Dop 
= \int_\Omega (\delta\mathfrak{F_1}/\delta u)\Dop(\delta\mathfrak{F_2}/\delta u)\;dx
\end{equation}
modulo boundary terms, 
with $\mathfrak{F}=\int_\Omega F\;dx$ 
denoting a functional on a domain $\Omega\subseteq\Rnum$,
where $F$ is an arbitrary differential function. 
A Poisson bracket \eqref{poissonbrack} satisfies 
\begin{equation}
\{\mathfrak{F_1},\mathfrak{F_2}\}_\Dop  + \{\mathfrak{F_2},\mathfrak{F_1}\}_\Dop  =0,
\quad
\{\{\mathfrak{F_1},\mathfrak{F_2}\}_\Dop,\mathfrak{F_3}\}_\Dop  + \text{ cyclic } =0
\end{equation}
which corresponds to the operator $\Dop$ being Hamiltonian. 
In particular, 
the Poisson bracket properties can be shown to be equivalent to the conditions \cite{Olv} 
\begin{gather}
\eta_1(\Dop \eta_2) + \eta_2(\Dop \eta_1) \equiv 0
\label{Dopskewprop}\\
\eta_1(\delta_{\Dop\eta_2}\Dop) \eta_3 + \text{ cyclic } \equiv 0
\label{Dopjacobiprop}
\end{gather}
holding for all differential functions $\eta_i$, 
where the notation ``$\equiv$'' denotes equality modulo a total $x$-derivative $D_x\theta$. 

Two Hamiltonian operators $\Dop_1$ and $\Dop_2$ are said to be a compatible pair if 
an arbitrary linear combination $c_1\Dop_1+c_2\Dop_2$ is again a Hamiltonian operator. 
This is equivalent to the condition \cite{Olv} 
\begin{equation}
\eta_1(\delta_{\Dop_1\eta_2}\Dop_2) \eta_3 + \eta_1(\delta_{\Dop_2\eta_2}\Dop_1) \eta_3 
+ \text{ cyclic } \equiv 0
\label{compatibleprop}
\end{equation}
holding for all differential functions $\eta_i$. 

Suppose the root equation in a hierarchy \eqref{integrhierarchy} of integrable evolution equations
has a bi-Hamiltonian structure
\begin{equation}\label{bihamileqn}
u_t = \Rop(u_x) 
= \Hop(\delta\mathfrak{H}/\delta u) 
= \Eop(\delta\mathfrak{E}/\delta u) 
\end{equation}
given by a compatible pair of Hamiltonian operators $\Hop$, $\Eop$,
and two Hamiltonian functionals $\mathfrak{H}$, $\mathfrak{E}$. 
Then Magri's theorem \cite{Mag}
shows that all of the higher-order evolution equations in the hierarchy 
inherit a bi-Hamiltonian structure,
and that the hereditary recursion operator for the hierarchy has the factorization 
\begin{equation}\label{EHRop}
\Rop = \Eop\Hop^{-1} . 
\end{equation}
There is another factorization of this operator 
\begin{equation}\label{HJRop}
\Rop = \Hop\Jop, 
\quad
\Jop= \Hop^{-1}\Eop\Hop^{-1}
\end{equation}
where $\Jop$ is a symplectic operator. 

The properties of a symplectic operator $\Jop$ have a natural formulation in terms of 
an associated symplectic 2-form defined by 
\begin{equation}\label{sympl2form}
\omega(\X_1,\X_2)_\Jop 
= \int_\Omega \eta_1\Jop(\eta_2)\;dx
\end{equation}
modulo boundary terms, 
with $\X=\eta\p_u$ being a vector field in evolutionary form, 
where $\eta$ is an arbitrary differential function. 
A symplectic 2-form \eqref{sympl2form} is skew and closed, 
\begin{equation}
\w(\X_1,\X_2)_\Jop + \w(\X_2,\X_1)_\Jop =0,
\quad
d\w(\X_1,\X_2,\X_3)_\Jop = \pr\X_1\w(\X_2,\X_3)_\Jop + \text{ cyclic } =0 . 
\end{equation}
These properties correspond to the operator $\Jop$ being symplectic. 
An equivalent characterization is that an operator $\Jop$ is symplectic iff
the formal inverse operator $\Jop^{-1}$ is Hamiltonian. 

A standard feature of the hereditary recursion operator \eqref{EHRop} is that 
it can be used to produce a hierarchy of mutually compatible Hamiltonian operators
\begin{equation}
\Hop^{(n)} = \Rop^n\Hop, 
\quad
n=0,1,2,\ldots
\end{equation}
starting from the pair of Hamiltonian operators 
$\Hop^{(0)} = \Hop$, $\Hop^{(1)} = \Eop$. 
Moreover, 
all of these Hamiltonian operators $\Hop^{(n)}$ are compatible with the Hamiltonian operator $\Jop^{-1}$. 

We will now show that each generalized negative flow equation \eqref{generalflow} 
possesses a bi-Hamiltonian structure. 
To begin, we write equation \eqref{generalflow} in the form 
\begin{equation}\label{heqn}
h_{(n)} = u_t,
\quad
\Hop^{(n-1)}\Jop(h_{(n)}) = \kappa_n h_{(n)},
\quad
n=1,2,\ldots
\end{equation}
with the use of the factorization \eqref{HJRop}. 
Then we let 
\begin{equation}\label{wJheqn}
w_{(n)} = \Jop(h_{(n)}) ,
\quad
n=1,2,\ldots
\end{equation}
so that 
\begin{equation}
u_t = \Jop^{-1}(w_{(n)}),
\quad
n=1,2,\ldots
\end{equation}
where 
\begin{equation}\label{weqn}
\Hop^{(n-1)}(w_{(n)}) = \kappa_n \Jop^{-1}(w_{(n)}),
\quad
n=1,2,\ldots
\end{equation}
is equivalent to equation \eqref{heqn}. 
We next write equation \eqref{weqn} in the form 
\begin{equation}\label{Dweqn}
\Dop_n(w_{(n)}) = 0,
\quad
n=1,2,\ldots
\end{equation}
in terms of the operator 
\begin{equation}\label{Dop}
\Dop_n = \Hop^{(n-1)} - \kappa_n \Jop^{-1} . 
\end{equation}
It is straightforward to see that $\Dop_n$ is a Hamiltonian operator,
since $\Jop^{-1}$ and $\Hop^{(n-1)}$ are a compatible Hamiltonian pair. 

Recall, a differential function $f$ depending on $u$ and $x$-derivatives of $u$ 
will have the form of an Euler-Lagrange expression $f=E_u(L)$ 
iff its Frechet derivative operator is self-adjoint, $f{}^\prime=f{}^{\prime*}$. 
This is equivalent to the Helmholtz condition 
$\eta_1\delta_{\eta_2} f=\eta_2\delta_{\eta_1} f$, 
holding modulo a total $x$-derivative $D_x\theta$, 
for arbitrary differential functions $\eta$. 
When $f|_{u=0}$ is non-singular, 
a Lagrangian function $L$ can be determined from $f$ 
by a homotopy integral $L= u \int_0^1 f|_{u=u_{(\lambda)}} d\lambda$, 
with $u_{(\lambda)} = \lambda u$.
More generally, 
a function $f$ satisfying a pseudo-differential operator equation $\Dop f=0$, 
in which the coefficients of $\Dop$ depend on $u$ and $x$-derivatives of $u$, 
will have the form $f=\delta L/\delta u$ iff $\eta_1\delta_{\eta_2} f=\eta_2\delta_{\eta_1} f$, 
holds modulo a total $x$-derivative $D_x\theta$, 
for arbitrary differential functions $\eta$. 

We will now state the main result and give its proof afterwards. 

\begin{thm}\label{biHamilnegflow}
Each generalized negative flow equation \eqref{generalflow} has
the bi-Hamiltonian structure
\begin{equation}\label{bihamilgeneralflow}
u_t = \Jop^{-1}(\delta\mathfrak{H}_{(n)}/\delta u) 
= \Dop_n(\delta\mathfrak{E}_{(n)}/\delta u),
\quad
n=1,2,\ldots
\end{equation}
where $\Dop_n=\Hop^{(n-1)} - \kappa_n \Jop^{-1}$ and $\Jop^{-1}$ 
are a compatible pair of Hamiltonian operators, 
and where 
$\mathfrak{H}_{(n)}=\int_\Omega H_{(n)}\;dx$ and $\mathfrak{E}_{(n)}=\int_\Omega E_{(n)}\;dx$ 
are Hamiltonian functionals 
with 
\begin{equation}\label{HEnegflow}
H_{(n)}= u \int_0^1 (\Jop(h_{(n)}))|_{u=u_{(\lambda)}} d\lambda,
\quad
E_{(n)}= u \int_0^1 (\Dop_n^{-1}(h_{(n)}))|_{u=u_{(\lambda)}} d\lambda,
\quad
u_{(\lambda)} = \lambda u
\end{equation}
and with $h_{(n)}$ defined implicitly in terms of $u$ 
by equations \eqref{wJheqn} and \eqref{weqn}. 
\end{thm}

The proof consists of showing that both 
$w_{(n)} = \Jop(h_{(n)})$ and $\tilde w_{(n)} = \Dop_n^{-1}(h_{(n)})$ 
satisfy the Helmholtz condition. 
Note that $w_{(n)} = \Jop\Dop_n(\tilde w_{(n)}) = (\Rop^*{}^{n-1} -\kappa_n)\tilde w_{(n)}$
from equation \eqref{Dop},
where $\Rop^*=\Jop\Hop$ is the adjoint recursion operator. 

To proceed with the proof, 
from equation \eqref{Dweqn} we have 
$0=\delta_{\eta_1}(\Dop_n w_{(n)}) = (\delta_{\eta_1}\Dop_n) w_{(n)} + \Dop_n(\delta_{\eta_1} w_{(n)})$
for an arbitrary differential function $\eta_1$. 
We multiply by another arbitrary differential function $\tilde\eta_2$ to get 
$\tilde\eta_2(\delta_{\eta_1}\Dop_n) w_{(n)} \equiv (\Dop_n\tilde\eta_2)\delta_{\eta_1} w_{(n)}$
after integration by parts using $\Dop_n^*=-\Dop_n$. 
Next, we write $\eta_i=\Dop_n\tilde\eta_i$, which gives 
$\tilde\eta_2(\delta_{\Dop_n\tilde\eta_1}\Dop_n) w_{(n)} \equiv \eta_2\delta_{\eta_1} w_{(n)}$.
By antisymmetrizing this equation in $\eta_1$ and $\eta_2$, we obtain
\begin{equation}
\eta_2 \delta_{\eta_1} w_{(n)} -\eta_1 \delta_{\eta_2} w_{(n)}
\equiv 
\tilde\eta_2(\delta_{\Dop_n\tilde\eta_1}\Dop_n) w_{(n)} 
-\tilde\eta_1(\delta_{\Dop_n\tilde\eta_2}\Dop_n) w_{(n)} 
\end{equation}
where the lefthand side will vanish iff $w_{(n)}$ satisfies the Helmholtz condition. 
The righthand side can be simplified by using the Hamiltonian properties 
\eqref{Dopskewprop}--\eqref{Dopjacobiprop} of $\Dop_n$ to get 
\begin{equation}
\tilde\eta_2(\delta_{\Dop_n\tilde\eta_1}\Dop_n) w_{(n)} 
-\tilde\eta_1(\delta_{\Dop_n\tilde\eta_2}\Dop_n) w_{(n)} 
\equiv 
\tilde\eta_1(\delta_{\Dop_n w_{(n)}}\Dop_n) \tilde\eta_2
\end{equation}
This expression vanishes due to equation \eqref{Dweqn}. 
Hence,  we have 
\begin{equation}
\eta_2 \delta_{\eta_1} w_{(n)} \equiv \eta_1 \delta_{\eta_2} w_{(n)}
\end{equation}
showing that $w_{(n)}$ satisfies the Helmholtz condition. 

To complete the proof, 
we now show by a similar kind of argument that 
\begin{equation}\label{wDinvH}
\tilde w_{(n)} = \Dop_n^{-1}(h_{(n)}) 
\end{equation}
satisfies the Helmholtz condition. 
Taking the Frechet derivative of this relation \eqref{wDinvH}, 
we have 
$\delta_{\eta_1}\tilde w_{(n)} = \delta_{\eta_1}(\Dop_n^{-1}h_{(n)}) 
=(\delta_{\eta_1}\Dop_n^{-1}) h_{(n)} + \Dop_n^{-1}(\delta_{\eta_1} h_{(n)})$
for an arbitrary differential function $\eta_1$. 
We multiply by another arbitrary differential function $\eta_2$,
and integrate by parts, 
yielding
\begin{equation}
\eta_2\delta_{\eta_1}\tilde w_{(n)} 
= (\Dop_n^{-1}\eta_2)(\delta_{\eta_1}\Dop_n) w_{(n)} - (\Dop_n^{-1}\eta_2) \delta_{\eta_1} h_{(n)}
\end{equation}
after  use of the identity 
$\delta_\eta\Dop^{-1}= -\Dop^{-1}(\delta_\eta\Dop)\Dop^{-1}$. 
Now we write $\eta_i=\Dop_n\tilde\eta_i$, and antisymmetrize in $\eta_1$ and $\eta_2$, 
which gives 
\begin{equation}\label{varweqn}
\eta_2\delta_{\eta_1}\tilde w_{(n)} - \eta_1\delta_{\eta_2}\tilde w_{(n)} 
= \tilde\eta_1\delta_{\Dop_n\tilde\eta_2} h_{(n)} - \tilde\eta_2\delta_{\Dop_n\tilde\eta_1} h_{(n)}
+ \tilde\eta_2(\delta_{\Dop_n\tilde\eta_1}\Dop_n)\tilde w_{(n)} -\tilde\eta_1(\delta_{\Dop_n\tilde\eta_2}\Dop_n)\tilde w_{(n)} . 
\end{equation}
The lefthand side of this equation 
will vanish iff $\tilde w_{(n)}$ satisfies the Helmholtz condition. 
On the righthand side, 
the last two terms can be simplified by using the Hamiltonian properties 
\eqref{Dopskewprop}--\eqref{Dopjacobiprop} of $\Dop_n$ to get 
\begin{equation}\label{varweqn-lastterms}
\tilde\eta_2(\delta_{\Dop_n\tilde\eta_1}\Dop_n)\tilde w_{(n)} 
-\tilde\eta_1(\delta_{\Dop_n\tilde\eta_2}\Dop_n)\tilde w_{(n)} 
\equiv 
-\tilde\eta_1(\delta_{\Dop_n\tilde w_{(n)}}\Dop_n) \tilde\eta_2
= -\tilde\eta_1(\delta_{h_{(n)}}\Dop_n) \tilde\eta_2 .
\end{equation}
Next, the first two terms on the righthand side of equation \eqref{varweqn}
can be expanded by using the relation $h_{(n)}=\Jop^{-1} w_{(n)}$,
which yields 
\begin{equation}\label{varweqn-firstterms}
\begin{aligned}
\tilde\eta_1\delta_{\Dop_n\tilde\eta_2} h_{(n)} - \tilde\eta_2\delta_{\Dop_n\tilde\eta_1} h_{(n)}
& = \tilde\eta_1(\delta_{\Dop_n\tilde\eta_2}\Jop^{-1}) w_{(n)} -\tilde\eta_2(\delta_{\Dop_n\tilde\eta_1}\Jop^{-1}) w_{(n)} 
\\&\qquad
+ \tilde\eta_1\Jop^{-1}(\delta_{\Dop_n\tilde\eta_2}w_{(n)}) -\tilde\eta_2\Jop^{-1}(\delta_{\Dop_n\tilde\eta_1}w_{(n)}) . 
\end{aligned}
\end{equation}
The first two terms on the righthand side of this equation 
can be simplified by using the property \eqref{compatibleprop}
that the Hamiltonian operators $\Dop_n$ and $\Jop^{-1}$ are compatible,
combined with the property that $\Jop^{-1}$ is skew. 
This gives 
\begin{equation}\label{varweqn-firstterms-simplified}
\begin{aligned}
\tilde\eta_1(\delta_{\Dop_n\tilde\eta_2}\Jop^{-1}) w_{(n)} -\tilde\eta_2(\delta_{\Dop_n\tilde\eta_1}\Jop^{-1}) w_{(n)} 
& \equiv
\tilde\eta_1(\delta_{\Dop_n w_{(n)} }\Jop^{-1}) \tilde\eta_2
+\tilde\eta_1(\delta_{\Jop^{-1}w_{(n)}}\Dop_n)\tilde\eta_2
\\&\qquad
-\tilde\eta_1(\delta_{\Jop^{-1}\tilde\eta_2}\Dop_n) w_{(n)} +\tilde\eta_2(\delta_{\Jop^{-1}\tilde\eta_1}\Dop_n) w_{(n)} . 
\end{aligned}
\end{equation}
The first term on the righthand side in equation \eqref{varweqn-firstterms-simplified}
vanishes due to the flow equation \eqref{Dweqn},
while the second term cancels the term \eqref{varweqn-lastterms}. 
Hence, after all of these steps, 
the righthand side of \eqref{varweqn} is given by combining 
the remaining two terms on the righthand side in equations 
\eqref{varweqn-firstterms-simplified} and \eqref{varweqn-firstterms},
which gives 
\begin{equation}
\begin{aligned}
& \tilde\eta_1\delta_{\Dop_n\tilde\eta_2} h_{(n)} - \tilde\eta_2\delta_{\Dop_n\tilde\eta_1} h_{(n)}
+ \tilde\eta_2(\delta_{\Dop_n\tilde\eta_1}\Dop_n)\tilde w_{(n)} -\tilde\eta_1(\delta_{\Dop_n\tilde\eta_2}\Dop_n)\tilde w_{(n)} 
\\& \equiv
\tilde\eta_1\Jop^{-1}(\delta_{\Dop_n\tilde\eta_2}w_{(n)}) -\tilde\eta_2\Jop^{-1}(\delta_{\Dop_n\tilde\eta_1}w_{(n)}) 
-\tilde\eta_1(\delta_{\Jop^{-1}\tilde\eta_2}\Dop_n) w_{(n)} +\tilde\eta_2(\delta_{\Jop^{-1}\tilde\eta_1}\Dop_n) w_{(n)} . 
\end{aligned}
\end{equation}
Finally, we write 
$\tilde\eta_i=\Jop\hat\eta_i$, 
so then equation \eqref{varweqn} becomes
\begin{equation}\label{varweqn-final}
\begin{aligned}
\eta_2\delta_{\eta_1}\tilde w_{(n)} - \eta_1\delta_{\eta_2}\tilde w_{(n)} 
& \equiv
\hat\eta_2(\delta_{\tilde\Rop\hat\eta_1}w_{(n)}) -\hat\eta_1(\delta_{\tilde\Rop\hat\eta_2}w_{(n)}) 
\\&\qquad
+\Jop\hat\eta_2(\delta_{\hat\eta_1}\Dop_n) w_{(n)} -\Jop\hat\eta_1(\delta_{\hat\eta_2}\Dop_n) w_{(n)} 
\end{aligned}
\end{equation}
after integration by parts, using $\Jop^*=-\Jop$, 
where $\tilde\Rop =\Dop_n\Jop$. 
The last two terms on the righthand side of equation \eqref{varweqn-final} 
can be converted into the form 
\begin{equation}
\begin{aligned}
\Jop\hat\eta_2(\delta_{\hat\eta_1}\Dop_n) w_{(n)} -\Jop\hat\eta_1(\delta_{\hat\eta_2}\Dop_n) w_{(n)} 
& = 
-(\Jop\hat\eta_2)\Dop_n(\delta_{\hat\eta_1} w_{(n)})  + (\Jop\hat\eta_1)\Dop_n(\delta_{\hat\eta_2} w_{(n)})
\\
&
\equiv 
(\tilde\Rop\hat\eta_2)\delta_{\hat\eta_1} w_{(n)}  -(\tilde\Rop\hat\eta_1)\delta_{\hat\eta_2} w_{(n)}
\end{aligned}
\end{equation}
by using equation \eqref{Dweqn} followed by integration by parts using $\Dop_n^*=-\Dop_n$. 
As a result, 
equation \eqref{varweqn-final} becomes
\begin{equation}
\eta_2\delta_{\eta_1}\tilde w_{(n)} - \eta_1\delta_{\eta_2}\tilde w_{(n)} 
\equiv
\hat\eta_2(\delta_{\tilde\Rop\hat\eta_1}w_{(n)}) 
-(\tilde\Rop\hat\eta_1)\delta_{\hat\eta_2} w_{(n)}
+(\tilde\Rop\hat\eta_2)\delta_{\hat\eta_1} w_{(n)}  
-\hat\eta_1(\delta_{\tilde\Rop\hat\eta_2}w_{(n)}) . 
\end{equation}
The two groups of terms on the righthand side separately vanish 
because $w_{(n)}$ satisfies the Helmholtz condition. 
Hence, we have 
\begin{equation}
\eta_2\delta_{\eta_1}\tilde w_{(n)} = \eta_1\delta_{\eta_2}\tilde w_{(n)} 
\end{equation}
showing that $\tilde w_{(n)}$ satisfies the Helmholtz condition. 
This completes the proof. 

We remark that when the Hamiltonian operators $\Hop^{(n-1)}$ and $\Jop^{-1}$ 
possess a scaling symmetry
then the negative flow equation \eqref{bihamilgeneralflow} with $\kappa_n=0$ 
will also possess a scaling symmetry, 
$x\rightarrow\lambda x$, $u\rightarrow\lambda^{-q} u$, $t\rightarrow\lambda^{-m_n} t$. 
In this case, the Hamiltonians \eqref{HEnegflow} can be obtained from $h_{(n)}$ 
by an algebraic formula \cite{Anc03} using the scaling generator 
$\X=(-qu-xu_x+m_ntu_t)\p_u$ 
instead of the homotopy integral formula. 
In particular, 
let $s_n$ be the scaling weight of $w_{(n)}=\Jop(h_{(n)}) =\delta H_{(n)}/\delta u$
and consider the action of the scaling generator on $H_{(n)}$. 
Note that here both $w_{(n)}$ and $H_{(n)}$ are regarded as expressions 
depending on $u$ as defined by equations \eqref{wJheqn} and \eqref{weqn}. 
Consequently, for the first Hamiltonian $H_{(n)}$, we have 
\begin{equation}\label{varH1}
\X H_{(n)} = (s_n-q)H_{(n)} -xD_x H_{(n)} \equiv (s_n-q+1)H_{(n)}
\end{equation}
after integration by parts,
and we also have the variational relation 
$\X H_{(n)} \equiv w_{(n)} \X u = -(qu+xu_x) w_{(n)}$. 
The term involving $x$ in this relation can be expressed as 
$-xu_x w_{(n)} \equiv D_x^{-1}(u_x w_{(n)})$ 
using integration by parts,
which yields
\begin{equation}\label{varH2}
\X H_{(n)} \equiv -qu w_{(n)} +D_x^{-1}(u_x w_{(n)})  . 
\end{equation}
Equating the two expressions \eqref{varH1} and \eqref{varH2}, 
we obtain 
\begin{equation}\label{algHnegflow}
(s_n-q+1)H_{(n)} \equiv -qu w_{(n)} +D_x^{-1}(u_x w_{(n)})
\end{equation}
which yields $H_{(n)}$ provided that $s_n\neq q-1$. 
A similar algebraic formula holds for the second Hamiltonian $E_{(n)}$,
\begin{equation}\label{algEnegflow}
(\tilde s_n-q+1)H_{(n)} \equiv -qu \tilde w_{(n)} +D_x^{-1}(u_x\tilde w_{(n)})
\end{equation}
starting from $\tilde w_{(n)}=\Dop_n^{-1}(h_{(n)}) =\delta E_{(n)}/\delta u$, 
where $\tilde s_n$ denotes its scaling weight.

\subsection{First integrals of negative flows}

The bi-Hamiltonian structure \eqref{bihamilgeneralflow} for negative flows 
has an interesting consequence that each negative flow equation \eqref{generalflow} 
can be integrated at least once to produce an equivalent equation of lower differential order. 

\begin{prop}\label{firstintegral}
Each generalized negative flow equation \eqref{generalflow} 
with a bi-Hamiltonian structure \eqref{bihamilgeneralflow} 
satisfies 
\begin{equation}\label{generalflowDxeqn}
D_x( \Phi^x_{\Hop^{(n-1)}}(w_{(n)},w_{(n)}) -\kappa_n\Phi^x_\Jop(h_{(n)},h_{(n)}) )=0
\end{equation}
and 
\begin{equation}\label{generalflowDxeqn2}
D_x( \Phi^x_{\Hop^{(n-1)}}(w_{(n)},\tilde w_{(n)}) -\Phi^x_\Jop(h_{(n)},\tilde w_{(n)}) )=0
\end{equation}
where $h_{(n)}$ and $w_{(n)}$ are defined by equations \eqref{wJheqn}-- \eqref{weqn},
and where $\tilde w_{(n)}$ is defined by $\Dop_n(\tilde w_{(n)})=h_{(n)}$.
A conformal scaling of $t$ can be chosen to put 
either
\begin{equation}\label{generalflowconformalscaling}
\Phi^x_{\Hop^{(n-1)}}(w_{(n)},w_{(n)}) -\kappa_n\Phi^x_\Jop(h_{(n)},h_{(n)}) =c_n = \const
\end{equation}
or
\begin{equation}\label{generalflowconformalscaling2}
\Phi^x_{\Hop^{(n-1)}}(\Jop h_{(n)},\tilde w_{(n)}) -\Phi^x_\Jop(h_{(n)},\Hop\tilde w_{(n)}) =\tilde c_n = \const . 
\end{equation}
\end{prop}

The proof uses only the skew property of the Hamiltonian operators $\Hop$ and $\Jop$
combined with the adjoint identity \eqref{adjointid}. 
Starting from the formulation  \eqref{weqn} for the generalized negative flow equation \eqref{generalflow}, 
we multiply by $w_{(n)}$ to get 
\begin{equation}\label{wmultreqn}
0= ( \Hop^{(n-1)}(w_{(n)}) - \kappa_n \Jop^{-1}(w_{(n)}) )w_{(n)} . 
\end{equation}
The first term in this equation is a total $x$-derivative 
$w_{(n)} \Hop^{(n-1)}(w_{(n)})  = D_x\Phi^x_{\Hop^{(n-1)}}(w_{(n)},w_{(n)})$
by applying the adjoint identity to $\Hop^{(n-1)}{}^*=-\Hop^{(n-1)}$. 
Next, the second term in equation  \eqref{wmultreqn}
can be expressed as $w_{(n)}\Jop^{-1}(w_{(n)}) = h_{(n)}\Jop(h_{(n)})$,
which is a total $x$-derivative 
$h_{(n)}\Jop(h_{(n)}) = D_x\Phi^x_{\Jop}(h_{(n)},h_{(n)})$
by applying the adjoint identity to $\Jop^*=-\Jop$. 
Hence, equation  \eqref{wmultreqn} reduces to a total $x$-derivative \eqref{generalflowDxeqn}. 
The derivation of the equation \eqref{generalflowDxeqn2} is similar. 
We consider the product 
\begin{equation}\label{hweqn}
w_{(n)}(h_{(n)}- \Dop_n(\tilde w_{(n)}))=0 .
\end{equation}
In this equation, the first term is a total $x$-derivative 
$w_{(n)}h_{(n)} = h_{(n)}\Jop(h_{(n)}) = D_x\Phi^x_{\Jop}(h_{(n)},h_{(n)})$
while the second term can be expressed as 
$w_{(n)} \Dop_n(\tilde w_{(n)}) = w_{(n)} \Hop(\tilde w_{(n)}) -\kappa_n w_{(n)} \Jop^{-1}(\tilde w_{(n)}) 
= D_x\Phi^x_{\Hop}(w_{(n)},\tilde w_{(n)})  +\kappa_n (\tilde w_{(n)} h_{(n)} - \Jop(h_{(n)}) \Jop^{-1}(\tilde w_{(n)}))$
after use of the adjoint identity applied to $\Hop^*=-\Hop$
followed by the relation $w_{(n)} = \Jop(h_{(n)})$. 
Then applying the adjoint identity to $\Jop^*=-\Jop$, we have 
$\tilde w_{(n)} h_{(n)} - \Jop(h_{(n)}) \Jop^{-1}(\tilde w_{(n)}) 
= -D_x\Phi^x_{\Jop}(h_{(n)},\Jop^{-1}\tilde w_{(n)})$. 
Substituting these expressions back into equation \eqref{hweqn},
we obtain 
\begin{equation}
0=D_x( \Phi^x_{\Jop}(h_{(n)},h_{(n)}+\kappa_n \Jop^{-1}\tilde w_{(n)}) - \Phi^x_{\Hop}(w_{(n)},\tilde w_{(n)}) )
\end{equation}
which reduces to equation \eqref{generalflowDxeqn2} 
since $w_{(n)} = \Jop(h_{(n)})$ and $h_{(n)}=\Hop(\tilde w_{(n)})-\kappa_n\Jop^{-1}(\tilde w_{(n)})$. 

To complete the proof, we first integrate equation \eqref{generalflowDxeqn}
with respect to $x$ to obtain 
$\Phi^x_{\Hop^{(n-1)}}(w_{(n)},w_{(n)}) -\kappa_n\Phi^x_\Jop(h_{(n)},h_{(n)}) =f(t)$. 
We next observe that this expression is bilinear in $h_{(n)}=u_t$
since $w_{(n)}=\Jop(h_{(n)}) = \Jop(u_t)$. 
Hence we can use a change of variable $t\rightarrow \tilde t(t)$ 
given by $(d\tilde t/dt)^2 = |f(t)|$ to put $f(t)=\pm 1 =\const$,
which leaves the flow equation \eqref{generalflow} unchanged up to 
a conformal scaling factor. 
This yields equation \eqref{generalflowconformalscaling}. 
Similar steps applied to equation \eqref{generalflowDxeqn2} 
give equation \eqref{generalflowconformalscaling2}.

\subsection{Conservation laws of negative flows}

When an evolution equation on a spatial domain $\Omega\subseteq\Rnum$ 
has a Hamiltonian structure 
$u_t= \Hop(\delta\mathfrak{H}/\delta u)$, 
the Hamiltonian functional $\mathfrak{H}=\int_\Omega H(t,x,u,u_x,u_{xx},\ldots)dx$ 
yields a global conservation law if $H_t=0$. 
The derivation relies on the adjoint identity \eqref{adjointid} applied to the Hamiltonian operator $\Hop$, 
and on the variational identity \eqref{varid} applied to $D_tH=H{}^\prime(u_t)$.
Specifically,
first using the variational identity we combine  
$H{}^\prime(u_t) = u_tE_u(H) + D_x\Phi^x(H,u_t)$
with $u_t=\Hop(E_u(H))$ to get 
$H{}^\prime(u_t) = E_u(H)\Hop(E_u(H))  + D_t\Phi^t(H,u_t) + D_x\Phi^x(H,u_t)$. 
Next we use $E_u(H)\Hop(E_u(H))= D_x\Phi_\Hop^x(H,H)$ 
which holds by the skew property for $\Hop$ together with the adjoint identity. 
As result, we obtain 
$D_t H = D_x\Psi$ 
where $\Psi=\Phi^x(H,u_t) +\Phi_\Hop^x(H,H)$. 
Finally, we integrate this divergence equation over $\Omega$, 
yielding the global conservation law
\begin{equation}
\frac{d}{dt}\int_\Omega H\; dx = \Psi\big|_{\p\Omega}
\end{equation}
holding for all solutions $u(t,x)$ of the evolution equation 
$u_t= \Hop(\delta\mathfrak{H}/\delta u)$. 
The local form of this conservation law can be expressed as a divergence identity 
\begin{equation}\label{conslawid}
D_tH -D_x\Psi = (u_t-\Hop(\delta\mathfrak{H}/\delta u))Q,
\quad
Q=E_u(H)
\end{equation}
in which $H$ is the conserved density, $-\Psi$ is the flux, 
and $Q$ is the multiplier. 
In particular, 
$Q$ satisfies the adjoint of the determining equation for evolutionary symmetries $\X=\eta\p_u$,
which can be shown by taking the variational derivative of the divergence identity \eqref{conslawid} 
and restricting the resulting equation to solutions $u(t,x)$ of $u_t= \Hop(\delta\mathfrak{H}/\delta u)$. 
Solutions of the adjoint symmetry determining equation are commonly called 
cosymmetries or adjoint-symmetries. 

Now consider a hierarchy \eqref{integrhierarchy} of integrable evolution equations
having a Hamiltonian structure
\begin{equation}\label{hamilhierarchy}
u_t = \Rop^k(u_x) 
= \Hop(\delta\mathfrak{H}^{(k)}/\delta u), 
\quad
k=0,1,2,\ldots
\end{equation}
with time-independent Hamiltonian functionals 
\begin{equation}\label{Hhierarchy}
\mathfrak{H}^{(k)}=\int_\Omega H^{(k)}(x,u,u_x,u_{xx},\ldots)dx,
\quad
k=0,1,2,\ldots . 
\end{equation}
It is well-known that each Hamiltonian functional yields a global conservation law
for the root evolution equation $u_t=u_x$ in the hierarchy. 
We will give a simple proof of this result, 
which we will extend to all negative flow equations \eqref{generalflow} 
associated to the hierarchy. 

\begin{prop}\label{Hamilhierarchy}
In a hierarchy \eqref{hamilhierarchy} of Hamiltonian evolution equations, 
each Hamiltonian gradient $Q^{(k)}= E_u(H^{(k)})$, $k=0,1,2,\ldots$ is a multiplier
that yields a conservation law identity \eqref{conslawid} with $H=H^{(k)}$ 
holding for the root equation $u_t = u_x = \Hop(\delta\mathfrak{H}^{(0)}/\delta u)$.
\end{prop}

The proof uses the factorization $\Rop=\Hop\Jop$ where $\Jop$ is a symplectic operator. 
From the form of the evolution equations in the hierarchy, 
we have 
$\Rop^k(u_x) = \Hop(Q^{(k)})$
since $\delta\mathfrak{H}^{(k)}/\delta u = E_u(H^{(k)})$.
This yields 
\begin{equation}\label{Qid}
Q^{(k)}=\Hop^{-1}\Rop^k(u_x)=\Rop^*{}^{k-1}\Jop(u_x)
\end{equation}
where $\Rop^*=\Jop\Hop$ is the adjoint recursion operator. 
In particular, the gradient of the first Hamiltonian $H^{(0)}$ 
gives the multiplier $Q^{(0)}=\Hop^{-1}(u_x)$. 
Now, the variational identity \eqref{varid} shows that
\begin{equation}\label{Qutid}
Q^{(k)}u_t = D_tH^{(k)} - D_x\Phi^x(H^{(k)},u_t)
\end{equation}
since $H^{(k)}_t=0$. 
Next, the relation \eqref{Qid} yields
$Q^{(k)}\Rop(u_x) = (\Rop^*{}^{k-1}\Jop(u_x))\Hop\Jop(u_x)$
which can be expressed in a symmetrical form 
after repeated integration by parts 
followed by use of the commutation relation $\Rop\Hop=\Hop\Rop^*$. 
When $k$ is odd, 
we obtain 
\begin{equation}\label{QkR-evenk}
Q^{(k)}\Rop(u_x) \equiv 
(\Rop^*{}^{(k-1)/2}\Jop u_x)(\Hop(-\Rop^*)^{(k-1)/2}\Jop u_x)
= (-1)^{(k-1)/2}Q^{((k+1)/2)}\Hop Q^{((k+1)/2)} 
\end{equation}
whereas when $k$ is even, we get
\begin{equation}\label{QkR-oddk}
Q^{(k)}\Rop(u_x) \equiv 
\Rop^*(\Rop^*{}^{(k-2)/2}\Jop u_x)(\Hop(-\Rop^*)^{(k-2)/2}\Jop u_x)
= (-1)^{(k-2)/2}(\Jop\Hop Q^{(k/2)})\Hop Q^{(k/2)} 
\end{equation}
where the notation ``$\equiv$'' denotes equality modulo a total $x$-derivative $D_x\theta$. 
Finally, by using the skew property of $\Hop$ and $\Jop$,
along with the adjoint identity \eqref{adjointid}, 
we see that the righthand sides of equations \eqref{QkR-evenk} and \eqref{QkR-oddk}
are total $x$-derivatives, 
$Q^{((k+1)/2)}\Hop Q^{((k+1)/2)} =D_x\Psi^x_\Hop(Q^{((k+1)/2)},Q^{((k+1)/2)})$
and 
$(\Hop Q^{(k/2)})\Jop(\Hop Q^{(k/2)}) =D_x\Psi^x_\Jop(\Hop Q^{(k+1)/2},\Hop Q^{((k+1)/2)})$. 
Hence, we have established 
$(u_t -\Rop(u_x))Q^{(k)} \equiv D_t H^{(k)}$,
which completes the proof. 

It follows from \propref{Hamilhierarchy} that the root equation in the hierarchy \eqref{hamilhierarchy}
possesses infinitely many global conservation laws
\begin{equation}
\frac{d}{dt}\int_\Omega H^{(k)} \; dx = \Psi^{(k)} \big|_{\p\Omega},
\quad
k=0,1,2,\ldots . 
\end{equation}

We now state and prove a similar result for all negative flow equations \eqref{generalflow}.  

\begin{thm}\label{conslawsnegflow}
Each generalized negative flow equation \eqref{generalflow} 
associated to a hierarchy of Hamiltonian evolution equations\eqref{hamilhierarchy} 
possesses infinitely many global conservation laws
\begin{equation}\label{conslawhierarchy}
\frac{d}{dt}\int_\Omega \tilde{H}^{(k)}\; dx = \tilde\Psi^{(k)} \big|_{\p\Omega},
\quad
\tilde{H}^{(k)}=H^{(k+n)} -\kappa_n H^{(k)},
\quad
k=0,1,2,\ldots
\end{equation}
where $H^{(l)}$ is the time-independent Hamiltonian for the $l^\text{th}$ 
evolution equation. 
\end{thm}

The proof consists of showing that $Q^{(k)}= E_u(H^{(k)})$ is a multiplier 
producing a local conservation law for each negative flow equation. 
Consider $(\Rop^n(u_t) - \kappa_n u_t)Q^{(k)}$. 
The second term can be expressed as $Q^{(k)}u_t \equiv D_t H^{(k)}$
by \propref{Hamilhierarchy}. 
Similarly, the first term becomes
$Q^{(k)} \Rop^n(u_t) \equiv (\Rop^*{}^nQ^{(k)})u_t = Q^{(k+n)}u_t \equiv D_t H^{(k+n)}$
after repeated integration by parts followed by use of \propref{Hamilhierarchy}. 
This establishes $0=(\Rop^n(u_t) - \kappa_n u_t)Q^{(k)} \equiv D_t(H^{(k+n)} -\kappa_n H^{(k)})$,
which completes the proof.

\section{Examples of generalized -$1$ flows}
\label{scalarevoleqns}

The root equation in each of the six hierarchies of 
integrable semilinear positive-weight polynomial scalar evolution equations 
is listed in Table~\ref{rootintegreqns},
where $q$ denotes the scaling weight with respect to the group of scaling transformations
\begin{equation}
x\rightarrow\lambda x,
\quad
t\rightarrow\lambda^m t,
\quad
u\rightarrow\lambda^{-q} u
\end{equation}
(with group parameter $\lambda\neq 0$). 
Note hereafter we use the notation $u_{kx}=\p_x^k u$, $k=1,2,\ldots$. 

\begin{table}[ht!]
\begin{center}
\begin{tabular}{ | l | l | l | }
\hline
& Integrable evolution equation & $(m,q)$  \\ \hline
Burgers & $u_t=u_{2x}+uu_{x}$ & $(2,1)$  \\ \hline
KdV & $u_t=u_{3x}+uu_{x}$ & $(3,2)$  \\ \hline
mKdV & $u_t=u_{3x}+\tfrac{3}{2}u^2u_{x}$ & $(3,1)$  \\ \hline
Sawada-Kotera & $u_t=u_{5x}+5uu_{3x}+5u_{x}u_{2x}+5u^2u_{x}$ & $(5,2)$  \\ \hline
Kaup-Kupershmidt & $u_t=u_{5x}+5uu_{3x}+\tfrac{25}{2}u_{x}u_{2x}+5u^2u_{x}$ & $(5,2)$  \\ \hline
Kupershmidt & $u_t=u_{5x}+5(u_{x}-u^2)u_{3x}+5u_{2x}^2-20uu_{x}u_{2x}-5u_{x}^3+5u^4u_{x}$ & $(5,1)$  \\ \hline
\end{tabular}
\end{center}
\caption{Root equations in integrable hierarchies}
\label{rootintegreqns}
\end{table}

In each of these six hierarchies, 
the generalized $-1$ flow equation has the form 
\begin{equation}\label{general-1flow}
\Rop(u_t) =\kappa u_t,
\quad
\kappa=\const
\end{equation}
where $\Rop$ is the recursion operator for the hierarchy. 
Only the recursion operator for the Burgers hierarchy 
does not possess a factorization \eqref{HJRop} 
given by a Hamiltonian operator $\Hop$ and a compatible symplectic operator $\Jop$. 
A convenient summary of all these operators is provided in \Ref{Wan}. 

We now work out the explicit form and properties of the first generalized negative flow in each hierarchy. 
We will start with the Burgers hierarchy, 
which has no Hamiltonian structure. 
Next we will consider the mKdV and Kupershmidt hierarchies, 
which share both a scaling symmetry and a Hamiltonian operator. 
Last we will consider the KdV, Sawada-Kotera, and Kaup-Kupershmidt hierarchies,
which share a scaling symmetry and have similar first integrals.

\subsection{Burgers generalized -1 flow}

The recursion operator for the Burgers hierarchy is given by 
\begin{equation}
\Rop=D_x+\tfrac{1}{2}D_x(uD_x^{-1}) . 
\end{equation}
We can write the generalized $-1$ flow equation \eqref{general-1flow} 
in the Burgers hierarchy as a PDE system 
\begin{subequations}\label{burg-1flowsys}
\begin{gather}
u_t=h,
\label{burgh}\\
\Rop(h) = D_xh +\tfrac{1}{2}uh +\tfrac{1}{2}u_xh_1 = \kappa h,
\label{burgRh}\\
D_xh_1=h 
\label{burgh1}
\end{gather}
\end{subequations}
involving $u$, a negative-flow variable $h$, and a potential variable $h_1$. 
This system can be expressed equivalently in the form of a single non-evolutionary 
equation as follows. 

First, we solve equation \eqref{burgRh} for $h_1$, 
which yields
\begin{equation}\label{burgh1expr}
h_1=\frac{(2\kappa-u)h-2D_xh}{u_x} . 
\end{equation}
Next, we substitute $h_1$ into equation \eqref{burgh1}, 
giving 
\begin{equation}
2D_x^2h +(u-2\kappa)D_x h-u_x^{-1}u_{2x}(2D_x h +(u-2\kappa)h) +2u_x h =0
\end{equation}
after the terms have been expanded out. 
Finally, we substitute $h$ from equation \eqref{burgh}, 
which gives the equation 
\begin{equation}\label{burg-1flow}
u_{t2x}=(\kappa-\tfrac{1}{2}u+u_x^{-1}u_{2x})u_{tx}-(u_x+u_x^{-1}u_{2x}(\kappa-\tfrac{1}{2}u))u_t
\end{equation}
which is the generalized Burgers $-1$ flow. 

Although the Burgers $-1$ flow does not have a Hamiltonian structure, 
it does possess a first integral, 
which arises from the identity 
\begin{equation}
D_x\big( h+\tfrac{1}{2}uh_1 -\kappa h_1 \big)=0 
\end{equation}
holding for the PDE system \eqref{burg-1flowsys}. 
This identity yields 
$h+\tfrac{1}{2}uh_1 -\kappa h_1=c(t)$
where $c(t)$ can be scaled by a conformal change of variable 
$t\rightarrow \tilde t(t)$ 
so that $c=\tilde{c}=\const$ without loss of generality. 
From the resulting first integral 
\begin{equation}
h+\tfrac{1}{2}uh_1 -\kappa h_1=\tilde{c}
\end{equation}
we can express 
\begin{equation}
h_1=\frac{2(h-\tilde{c})}{2\kappa-u} . 
\end{equation}
Then, substituting this expression for $h_1$ along with $h=u_t$ into equation \eqref{burgh1expr}, 
we obtain the lower order non-evolutionary equation 
\begin{equation}\label{burg-1flowintegrated}
u_{tx}=
(\kappa-\tfrac{1}{2}u)u_t
+\frac{u_x(u_t-\tilde{c})}{u-2\kappa}
\end{equation}
which is equivalent to the generalized Burgers $-1$ flow equation \eqref{burg-1flow}
up to a conformal scaling of $t$. 

This $-1$ flow equation has three main properties. 
Firstly, 
it is invariant under the scaling transformation 
$x\rightarrow\lambda x$, $t\rightarrow\lambda^{-1} t$, $u\rightarrow\lambda^{-1} u$, 
$\kappa\rightarrow\lambda^{-1}\kappa$. 
Secondly, 
it possesses a hierarchy of symmetries \eqref{symmhierarchy} 
from \thmref{symmsnegflow},
where the root symmetry is an $x$-translation $\X^{(0)}=u_x\p_u$
and the first higher-order symmetry 
\begin{equation}
\X^{(1)}=(u_{2x} +uu_x)\p_u
\end{equation}
corresponds to the Burgers flow $u_t= u_{2x} +uu_x$. 
Thirdly, it can be mapped into a linear equation by a Cole-Hopf transformation
$u=2v_x/v$, which yields
\begin{equation}
v_{tx}-\kappa v_t = \tfrac{1}{2}\tilde{c}v . 
\end{equation}
The special case $\tilde{c}=0$ of this linear equation has been obtained previously 
by a related method in \Ref{QiaCaoStr} 
where generalized $-n$ flows for Burgers equation are studied.

\subsection{mKdV generalized -1 flow}

The recursion operator for the mKdV hierarchy has the Hamiltonian factorization
\begin{equation}
\Rop=\Hop\Jop,
\quad
\Hop=D_x,
\quad
\Jop=D_x+uD_x^{-1}u.
\end{equation}
We can write the generalized $-1$ flow equation \eqref{general-1flow} 
in the mKdV hierarchy as a PDE system 
\begin{subequations}\label{mkdv-1flowsys}
\begin{gather}
u_t=h,
\label{mkdvh}\\
\Hop(w) = D_x w = \kappa h,
\label{mkdvHw}\\
w = \Jop(h) =D_xh+uh_1, 
\label{mkdvJh}\\
D_xh_1=uh 
\label{mkdvh1}
\end{gather}
\end{subequations}
involving $u$, a negative-flow variable $h$, and a potential variable $h_1$. 
This system can be expressed equivalently in the form of a single non-evolutionary 
equation as follows. 
First, we substitute $w$ from equation \eqref{mkdvJh} into equation \eqref{mkdvHw}
to get 
\begin{equation}
D_x^2h+u^2 h +u_x h_1 -\kappa h =0 . 
\end{equation}
Next, we solve this equation for 
\begin{equation}\label{mkdvh1expr}
h_1=\frac{(\kappa-u^2)h-D_x^2h}{u_x} 
\end{equation}
and substitute it into equation \eqref{mkdvh1}, 
which yields 
\begin{equation}
-D_x^3h +u_x^{-1}u_{2x} D_x^2h +(\kappa-u^2)D_xh -(3uu_x +u_x^{-1}u_{2x}(\kappa-u^2))h =0
\end{equation}
after the terms have been expanded out. 
Finally, we substitute $h$ from equation \eqref{mkdvh}, 
giving the equation 
\begin{equation}\label{mkdv-1flow}
u_{t3x}=u_x^{-1}u_{2x}u_{t2x}-3uu_xu_t+(\kappa-u^2)(u_{tx}-u_x^{-1}u_{2x}u_t) 
\end{equation}
which is the generalized mKdV $-1$ flow. 

From \propref{firstintegral}, 
it follows that this equation \eqref{mkdv-1flow} can be integrated once 
to obtain an equivalent equation of lower differential order, 
by use of the first integral \eqref{generalflowDxeqn}. 
The expression for the first integral is derived from the identities
\begin{equation}
w\Hop(w)=D_x\big(\tfrac{1}{2} w^2\big),
\quad
h\Jop(h)=h(D_xh+uh_1)=D_x\big(\tfrac{1}{2}(h^2+h_1^2)\big) , 
\end{equation}
which together yield 
\begin{equation}
D_x\big( (D_xh+uh_1)^2 -\kappa(h^2+h_1^2) \big)=0 
\end{equation}
holding for the PDE system \eqref{mkdv-1flowsys}. 
This gives the first integral 
$(D_xh+uh_1)^2 -\kappa(h^2+h_1^2)=c(t)$
where $c(t)$ can be scaled by a conformal change of variable 
$t\rightarrow \tilde t(t)$ 
so that $c=\tilde{c}=\const$ without loss of generality. 
Then we can use the resulting equation 
\begin{equation}
(D_xh+uh_1)^2 -\kappa(h^2+h_1^2)=\tilde{c}
\end{equation}
to express 
\begin{equation}
h_1=\frac{-uD_xh\pm\sqrt{(u^2-\kappa)(\kappa h^2+\tilde{c})+\kappa(D_xh)^2}}{u^2-\kappa} . 
\end{equation}
Substituting this expression for $h_1$ along with $h=u_t$ into equation \eqref{mkdvh1expr}, 
we obtain the non-evolutionary equation 
\begin{equation}\label{mkdv-1flowintegrated}
u_{t2x}=
(\kappa-u^2)u_t
+\frac{u_x}{u^2-\kappa}\Big( uu_{tx}\mp\sqrt{(u^2-\kappa)(\kappa u_t^2+\tilde{c})+\kappa u_{tx}^2} \Big) 
\end{equation}
which is equivalent to the generalized mKdV $-1$ flow equation \eqref{mkdv-1flow}
up to a conformal scaling of $t$. 

When $\kappa=0$, 
the generalized mKdV $-1$ flow equation \eqref{mkdv-1flowintegrated} 
reduces to the ordinary mKdV $-1$ flow which appears in \Ref{QiaStr}. 
In turn, the ordinary mKdV $-1$ flow 
reduces to $u_{t2x}=-u^2u_t +u^{-1}u_xu_{tx}$ when $\tilde{c}=0$. 
This reduced equation is equivalent to the sine-Gordon equation by the following steps. 
First, the reduced equation possesses an integrating factor $Q=u_{tx}/u^2$ 
which yields $(u_{t2x} +u^2u_t -u^{-1}u_xu_{tx})Q = D_x( u_t^2 + (u_{tx}/u)^2 )=0$. 
This implies $u_t^2 + (u_{tx}/u)^2 =\hat{c}^2=\const$ 
after a conformal change of variable $t\rightarrow \tilde t(t)$. 
Hence we obtain $u_{tx} =\pm u\sqrt{\hat{c}^2-u_t^2}$. 
This equation is well-known to be equivalent to the sine-Gordon equation 
$\theta_{tx}= \pm\hat{c}\sin\theta$
through the standard transformation $u=\pm\theta_x$, $u_t=\hat{c}\sin\theta$. 
There is an alternative derivation of this equivalence, which is useful to see. 
The reduced equation $u_{t2x}=-u^2u_t +u^{-1}u_xu_{tx}$ 
can be directly written as a PDE system 
\begin{equation}
u_t=h,
\quad
w = \Jop(h) =D_xh+uh_1 =0,
\quad
D_xh_1=uh . 
\label{SGflow}
\end{equation}
This system clearly satisfies $D_x(h^2+h_1{}^2)=0$ 
which yields the first integral $h^2+h_1{}^2=\hat{c}^2=\const$.
We can use this first integral to express $h_1= \pm\sqrt{\hat{c}^2-h^2}$
and substitute this expression along with $h=u_t$ into $D_xh=-uh_1$,
directly giving $u_{tx}= \mp u\sqrt{\hat{c}^2-u_t^2}$. 

Hence, the generalized mKdV $-1$ flow equation \eqref{mkdv-1flowintegrated} 
represents a two-parameter generalization of the sine-Gordon equation. 
We now state some main properties of this $-1$ flow. 

Firstly, 
the $-1$ flow equation \eqref{mkdv-1flow} is invariant
under the scaling transformation 
$x\rightarrow\lambda x$, $t\rightarrow\lambda^{-2} t$, $u\rightarrow\lambda^{-1} u$, 
$\kappa\rightarrow\lambda^{-2}\kappa$. 
Secondly, 
from \thmref{symmsnegflow}, 
the $-1$ flow equation \eqref{mkdv-1flow} possesses 
a hierarchy of symmetries \eqref{symmhierarchy}.
The root symmetry is an $x$-translation $\X^{(0)}=u_x\p_u$,
and the first higher-order symmetry 
\begin{equation}
\X^{(1)}=(u_{3x} +\tfrac{3}{2}u^2u_x)\p_u
\end{equation}
corresponds to the mKdV flow $u_t= u_{3x} +\tfrac{3}{2}u^2u_x$. 
Thirdly, 
from \thmref{conslawsnegflow}, 
the $-1$ flow equation \eqref{mkdv-1flow} also possesses 
a hierarchy of global conservation laws \eqref{conslawhierarchy} 
which are related to the conservation laws of the mKdV flow. 
It is well-known that the first four mKdV conservation laws are given by 
the Hamiltonian (conserved) densities 
$H^{(0)}=u$, 
$H^{(1)}=\tfrac{1}{2}u^2$, 
$H^{(2)}=-\tfrac{1}{2}u_x^2 + \tfrac{1}{8}u^4$, 
and $H^{(3)}=\tfrac{1}{2}(u_{2x}^2 -u^2u_x^2) + \tfrac{1}{6}u^6$. 
This yields the corresponding conserved densities
\begin{gather}
\tilde{H}^{(0)}=\tfrac{1}{2} u^2 -\kappa u
\\
\tilde{H}^{(1)}=-\tfrac{1}{2}( u_x^2 -\tfrac{1}{4}u^4 +\kappa u^2 )
\\
\tilde{H}^{(2)}=\tfrac{1}{2}( u_{2x}^2 -u^2u_x^2 + \tfrac{1}{3}u^6 +\kappa (u_x^2 -\tfrac{1}{4}u^4) )
\end{gather}
for the $-1$ flow equation \eqref{mkdv-1flow}. 
Fourthly, 
the $-1$ flow equation \eqref{mkdv-1flow} has a bi-Hamiltonian structure \eqref{bihamilgeneralflow} from \thmref{biHamilnegflow}. 
This structure can be written down in a simple, explicit way in the case $\kappa=0$, 
by considering the algebraic scaling formulas \eqref{algHnegflow} and \eqref{algEnegflow}. 
Note $q=1$, $s=0$, $\tilde s=2$ respectively are the scaling weights of $u$, $w$, $\tilde w$. 
For the first Hamiltonian density, 
we see $s-q+1=0$ implies that the algebraic scaling formula \eqref{algHnegflow} 
cannot be applied. 
However, since $D_xw=0$, 
we can immediately conclude that $H_{(1)} \equiv u w$ yields the Hamiltonian density
for the first Hamiltonian structure \eqref{bihamilgeneralflow}. 
For the second Hamiltonian density, 
the algebraic scaling formula is applicable since $\tilde s -q +1 \neq 0$. 
This yields 
$2E_{(1)} \equiv -u\tilde w +D_x^{-1}(u_x\tilde w)  \equiv -h_1$
after integration by parts, using equation \eqref{mkdvh1}. 
Hence $E_{(1)} \equiv -\tfrac{1}{2}h_1$ yields the Hamiltonian density
for the second Hamiltonian structure \eqref{bihamilgeneralflow}.

\subsection{Kupershmidt generalized $-1$ flow}

The recursion operator for the K hierarchy has the Hamiltonian factorization
\begin{equation}
\Rop=\Hop\Jop,
\quad
\Hop=D_x,
\quad
\Jop=\Jop_1+\Jop_2+\Jop_3,
\end{equation}
where
\begin{equation}
\begin{aligned}
\Jop_1=&
D_x^5+3(u_xD_x^3+D_x^3u_x)-3(u^2D_x^3+D_x^3u^2)-3(u^2u_xD_x+D_xu^2u_x),
\\
\Jop_2=&
-2(u_{3x}D_x+D_xu_{3x}) -2(uu_{2x}D_x+D_xuu_{2x})
\\&\qquad
+\tfrac{5}{2}(u_x^2D_x+D_xu_x^2)
+\tfrac{9}{2}(u^4D_x+D_xu^4),
\\
\Jop_3=&
-2(u_{4x}+5(u_x-u^2)u_{2x}-5uu_x^2+u^5)D_x^{-1}u
\\&\qquad
-2uD_x^{-1}(u_{4x}+5(u_x-u^2)u_{2x}-5uu_x^2+u^5).
\end{aligned}
\end{equation}
Note 
\begin{equation}
\Jop(h)=\A-2\B D_x^{-1}(uh)-2uD_x^{-1}(\B h),
\end{equation}
where
\begin{align}
& \begin{aligned}
\A=& 
D_x^5h+6(u_x-u^2)D_x^3h+9(u_{2x}-2uu_x)D_x^2h
\\&\qquad
+(5u_{3x}-22uu_{2x}-13u_x^2-6u^2u_x+9u^4)D_xh
\\&\qquad
+(u_{4x}-8uu_{3x}-3(5u_x+u^2)u_{2x}-6uu_x^2+18u^3u_x)h,
\end{aligned}
\\
& \B= u_{4x}+5(u_x-u^2)u_{2x}-5uu_x^2+u^5.
\end{align}
We can then write the generalized $-1$ flow equation \eqref{general-1flow} in the K hierarchy as a PDE system
\begin{subequations}\label{kup-1flowsys}
\begin{gather}
u_t=h,
\label{kuph}\\
\Hop(w)=D_xw=\kappa h,
\label{kupHw}\\
w=\Jop(h)=\A-2\B h_1-2uh_2,
\label{kupJh}\\
D_xh_1=uh,
\label{kuph1}\\
D_xh_2=\B h 
\label{kuph2}
\end{gather}
\end{subequations}
which involves $u$, a negative-flow variable $h$, and two potential variables $h_1,h_2$. 
This system is equivalent to a single non-evolutionary equation 
which is obtained by the following steps. 

We substitute $w$ from equation \eqref{kupJh} into equation\eqref{kupHw}, 
yielding 
\begin{equation}
D_x\A-4u\B h-2h_1D_x\B-2u_xh_2-\kappa h=0 . 
\end{equation}
We now solve this equation for 
\begin{equation}\label{kuph2expr}
h_2=\frac{D_x\A-(4u\B +\kappa)h-2h_1D_x\B}{2u_x}
\end{equation}
and substitute it into equation \eqref{kuph2}, 
which gives 
\begin{equation}\label{kuph1eqn}
2\F h_1+2\C+u_xD_x^2\A-u_{2x}D_x\A+\kappa\G=0,
\end{equation}
where
\begin{equation}
\begin{gathered}
\C=((2uu_{2x}-3u_x^2)\B -3uu_xD_x\B) h -2uu_x\B D_xh ,
\\
\F=u_{2x}D_x\B-u_xD_x^2\B,
\quad
\G=u_{2x}h-u_xD_xh.
\end{gathered}
\end{equation}
Solving equation \eqref{kuph1eqn} for $h_1$, we get 
\begin{equation}\label{kuph1expr}
h_1=\frac{u_{2x}D_x\A-u_xD_x^2\A-2\C-\kappa\G}{2\F}.
\end{equation}
We substitute this expression into equation \eqref{kuph1}, 
which yields 
\begin{equation}\label{kupeqn}
\begin{aligned}
&
u_x\F D_x^3\A-u_xD_x\F D_x^2\A+(u_{2x}D_x\F-u_{3x}\F)D_x\A
\\&\qquad
-2\C D_x\F+2\F D_x\C+2uh\F^2+\kappa(\F D_x\G-\G D_x\F)=0.
\end{aligned}
\end{equation}
The highest derivative term in this equation comes from the $D_x^3\A$ term.
If we isolate the highest derivative term in $\A$ by writing
\begin{equation}\label{kuphighestder}
\A=D_x^5h+\I, 
\end{equation}
then the highest derivative term in equation \eqref{kupeqn} will be $D_x^8h$.
Finally, 
by solving equation \eqref{kupeqn} for $D_x^8h$ and then substituting $h=u_t$, 
we obtain the generalized K $-1$ flow equation
\begin{equation}\label{kup-1flow} 
\begin{aligned}
u_{t8x}=&
\frac{(u_{t7x}+D_x^2\I)D_x\F}{\F}
-\frac{D_x\F (u_{2x}(u_{t6x}+D_x\I)-2\C-\kappa\G)}{u_x\F}
\\&\qquad
+\frac{u_{3x}(u_{t6x}+D_x\I)-2u\F u_t-\kappa D_x\G-2D_x\C}{u_x}
-D_x^3\I 
\end{aligned}
\end{equation}
with 
\begin{align}
& \begin{aligned}
\I=&
6(u_x-u^2)u_{t3x}+9(u_{2x}-2uu_x)u_{t2x}
\\&\qquad
+(5u_{3x}-22uu_{2x}-13u_x^2-6u^2u_x+9u^4)u_{tx}
\\&\qquad
+(u_{4x}-8uu_{3x}-3(5u_x+u^2)u_{2x}-6uu_x^2+18u^3u_x)u_t,
\end{aligned}
\label{kupI}
\\
& \begin{aligned}
\F = & 
-u_xu_{6x} +u_{2x}u_{5x} +5(u^2-u_x)u_xu_{4x} 
-5((u^2+2u_x)u_{2x}-6uu_x^2)u_{3x}
\\&\qquad
+5u_{2x}^3 +30u_x^3u_{2x} -20u^3u_x^3, 
\end{aligned}
\label{kupF}
\\
& \begin{aligned}
\C= &
(-3uu_xu_{5x} +(2uu_{2x}-3u_x^2)u_{4x} 
+15(u^2-u_x)uu_xu_{3x}
-5 u(u_x+2u^2)u_{2x}^2
\\&\qquad
+(2u^6+65u^2u_x^2 -15 u_x^3)u_{2x} 
+30 uu_x^4
-18u^5u_x^2)u_t 
\\&\qquad
-2uu_x(u_{4x}+5(u_x-u^2)u_{2x}-5uu_x^2+u^5)u_{tx}, 
\end{aligned}
\label{kupC}
\\
& 
\G=
u_{2x}u_t-u_xu_{tx}.
\label{kupG}
\end{align}

From \propref{firstintegral}, 
we can integrate the flow equation \eqref{kup-1flow} 
once to obtain an equivalent equation of lower differential order, 
by use of the first integral \eqref{generalflowDxeqn}. 
The expression for the first integral is derived from the identities
\begin{align}
& w\Hop(w)=D_x\big(\tfrac{1}{2} w^2\big),
\\
& \begin{aligned}
h\Jop_1(h)=&
hD_x^5h+3(u_xhD_x^3h+hD_x^3(u_xh))-3(u^2hD_x^3h+hD_x^3(u^2h))
\\&\qquad
-3(u^2u_xhD_xh+hD_x(u^2u_xh))
\\
=& 
D_x\big(hD_x^4h-D_xhD_x^3h+\tfrac{1}{2}(D_x^2h)+3D_x^2((u_x-u^2)h^2)
\\&\qquad
-9D_x((u_x-u^2)h)D_xh-3u^2u_xh^2\big),
\end{aligned}
\\
& \begin{aligned}
h\Jop_2(h)=&
\tfrac{5}{2}(u_x^2hD_xh+hD_x(u_x^2h))-2(u_{3x}hD_xh+hD_x(u_{3x}h))
\\&\qquad
-2(uu_{2x}hD_xh+hD_x(uu_{2x}h)) +\tfrac{9}{2}(u^4hD_xh+hD_x(u^4h))
\\
=&
D_x\big(\tfrac{5}{2}u_x^2h^2-2u_{3x}h^2+\tfrac{9}{2}u^4h^2-2uu_{2x}h^2\big),
\end{aligned}
\\
& h\Jop_3(h)= -2h\B h_1-2uh_2h=D_x\big(-2h_1h_2\big) . 
\end{align}
Together, these identities yield
\begin{equation}\label{kup-1flowid}
D_x\big( \tfrac{1}{2} (\A-2\B h_1-2uh_2)^2 -\kappa(\J -2h_1h_2)\big) 
=0
\end{equation}
where
\begin{equation}
\begin{aligned}
\J=&
hD_x^4h-D_xhD_x^3h+\tfrac{1}{2}(D_x^2h)^2+3D_x^2((u_x-u^2)h^2)-9D_x((u_x-u^2)h)D_xh
\\&\qquad
-(2u_{3x}+2uu_{2x}-\tfrac{5}{2}u_x^2 +3u^2u_x -\tfrac{9}{2}u^4)h^2.
\end{aligned}
\end{equation} 
This gives a first integral 
\begin{equation}\label{kup1stintegral}
\tfrac{1}{2}(\A-2\B h_1-2uh_2)^2-\kappa(\J-2h_1h_2)=c(t) 
\end{equation}
holding for the PDE system \eqref{kup-1flowsys}. 
By a conformal change of variable $t\rightarrow\tilde{t}$, 
we can put $c=\tilde{c}=\const$ without loss of generality. 
Then, we substitute $h_2$ from equation \eqref{kuph2expr} 
into this first integral \eqref{kup1stintegral}, 
which gives a quadratic equation in $h_1$, 
\begin{equation}\label{kuph1quadreqn}
\K h_1^2 + \M h_1 + \N =0
\end{equation}
where
\begin{align}
& \K= 4((uD_xB -u_xB)^2 -\kappa u_xD_xB), 
\\
& \M= 4(uD_xB -u_xB)(u_xA+u((4uB+\kappa) h-D_xA))+2\kappa u_x(D_xA -(4uB+\kappa)h),
\\
& \N = (u_x A+u((4uB+\kappa) h-D_xA))^2 -2(\kappa\J+\tilde c) u_x^2.
\end{align}
Solving equation \eqref{kuph1quadreqn}, 
and substituting $h_1$ into equation \eqref{kuph1expr}, 
we get 
\begin{equation}\label{kupeqnlowerord}
(u_{2x}D_x\A-u_xD_x^2\A-2\C-\kappa\G)\K =(-\M\pm\sqrt{\M^2-4\K\N})\F .
\end{equation}
The highest derivative term in this equation \eqref{kupeqnlowerord} comes from the $D_x^2A$ term. 
This gives, using equation \eqref{kuphighestder}, 
\begin{equation}
D_x^2\A=D_x^7h+D_x^2\I . 
\end{equation}
Hence, 
by solving equation \eqref{kupeqnlowerord} for $D_x^7h$ and then substituting $h=u_t$, 
we obtain the lower-order non-evolutionary equation 
\begin{equation}\label{kup-1flowintegrated}
u_{t7x}=
\frac{u_{2x}(u_{t6x}+D_x\I) -2\C-\kappa\G}{u_x} 
+\frac{(\M\mp\sqrt{\M^2-4\K\N})\F}{u_x\K} 
-D_x^2\I
\end{equation}
where
\begin{align}
&\begin{aligned}
\J=& 
u_tu_{t4x}-u_{tx}u_{t3x}+\tfrac{1}{2}u_{t2x}^2+6(u_x-u^2)u_tu_{t2x}+3(-u_x+u^2)u_{xt}^2
\\&\qquad
+3(u_{2x}-2uu_x)u_tu_{xt}+(u_{3x}-8uu_{2x}-\tfrac{7}{2}u_x^2-3u^2u_x+\tfrac{9}{2}u^4)u_t^2, 
\end{aligned}
\\
& \K=4(\K_1^2-\kappa\K_2), 
\\
& \M = 4\K_1\M_1 -2\kappa u_x\M_2, 
\\
& \N= \M_1^2 -2u_x^2(\kappa \J +\tilde{c}), 
\\
& \K_1=uu_{5x}-u_xu_{4x}+5u(u_x-u^2)u_{3x}+5uu_{2x}^2-5(3u^2+u_x)u_xu_{2x}+4u^5u_x, 
\\
& \K_2=u_x(u_{5x}+5(u_x-u^2)u_{3x}-5u_{2x}^2-20uu_xu_{2x}-5u_x^3+5u^4u_x), 
\\
&\begin{aligned}
\M_1 &= 
-uu_{t6x}+u_xu_{t5x}-6u(u_x-u^2)u_{t4x}+(-15uu_{2x}+24u^2u_x+6u_x^2)u_{t3x}
\\&\qquad
+(-14uu_{3x}+(40u^2+9u_x)u_{2x}+13uu_x^2+6u^3u_x-9u^5)u_{t2x}
\\&\qquad
+(-6uu_{4x}+5(u_x+6u^2)u_{3x}+(41uu_x+9u^3)u_{2x}-13u_x^3+12u^2u_x^2 
-45u^4u_x)u_{tx}
\\&\qquad
+(-uu_{5x}+(12u^2+u_x)u_{4x}+3(5uu_x+u^3)u_{3x}+15uu_{2x}^2
\\&\qquad
+(35u^2u_x-15u_x^2-38u^4)u_{2x}-56u^3u_x^2+4u^7+\kappa u)u_t, 
\end{aligned}
\\
&\begin{aligned}
\M_2 & = 
u_{t6x}+6(u_x-u^2)u_{t4x}+15(u_{2x}-2uu_x)u_{t3x}
\\&\qquad
+(14u_{3x}-40uu_{2x}-31u_x^2 -6u^2u_x+9u^4)u_{t2x}
\\&\qquad
+(6u_{4x}-30uu_{3x}-9(7u_x+u^2)u_{2x}-18uu_x^2+54u^3u_x)u_{tx}
\\&\qquad
+(u_{5x}-12uu_{4x} -(23u_x^2+3u^2)u_{3x} -15u_{2x}^2
-38u(u_x+u^2)u_{2x} 
\\&\qquad
-6u_x^3+74u^2u_x^2-4u^6 -\kappa)u_t, 
\end{aligned}
\end{align}
and where $I$, $F$, $C$, $G$ are given by expressions \eqref{kupI}--\eqref{kupG}.
This equation \eqref{kup-1flowintegrated} is equivalent to the generalized K $-1$ flow equation \eqref{kup-1flow} 
up to a conformal scaling of $t$. 

Finally, we state some main properties of this $-1$ flow. 

Firstly, 
the $-1$ flow equation \eqref{kup-1flow} is invariant
under the scaling transformation 
$x\rightarrow\lambda x$, $t\rightarrow\lambda^{-6} t$, $u\rightarrow\lambda^{-1} u$, 
$\kappa\rightarrow\lambda^{-2}\kappa$. 
Secondly, 
from \thmref{symmsnegflow}, 
the $-1$ flow equation \eqref{kup-1flow} possesses 
a hierarchy of symmetries \eqref{symmhierarchy}.
The root symmetry is an $x$-translation $\X^{(0)}=u_x\p_u$,
and the first higher-order symmetry 
\begin{equation}
\X^{(1)}=(u_{5x}+5(u_{x}-u^2)u_{3x}+5u_{2x}^2-20uu_{x}u_{2x}-5u_{x}^3+5u^4u_{x})\p_u
\end{equation}
corresponds to the Kupershmidt flow 
$u_t=u_{5x}+5(u_{x}-u^2)u_{3x}+5u_{2x}^2-20uu_{x}u_{2x}-5u_{x}^3+5u^4u_{x}$. 
Thirdly, 
from \thmref{conslawsnegflow}, 
the $-1$ flow equation \eqref{kup-1flow} also possesses 
a hierarchy of global conservation laws \eqref{conslawhierarchy} 
which are related to the conservation laws of the Kupershmidt flow. 
The first three of the Kupershmidt conservation laws \cite{FucOev}
are given by the Hamiltonian (conserved) densities 
$H^{(0)}=u$, 
$H^{(1)}=\tfrac{1}{2}u^2$, 
and $H^{(2)}=\tfrac{1}{2}u_{2x}^2 -\tfrac{5}{6}u_x^3 +\tfrac{5}{2}u^2u_x^2 + \tfrac{1}{6}u^6$. 
This yields the corresponding conserved densities
\begin{gather}
\tilde{H}^{(0)}=\tfrac{1}{2} u^2 -\kappa u 
\\
\tilde{H}^{(1)}=\tfrac{1}{2}( u_{2x}^2 -\tfrac{5}{3}u_x^2 +5u^2u_x^2 + \tfrac{1}{3}u^6 -\kappa u^2 )
\end{gather}
for the $-1$ flow equation \eqref{kup-1flow}. 
Fourthly, 
from \thmref{biHamilnegflow}, 
the $-1$ flow equation \eqref{kup-1flow} has a bi-Hamiltonian structure \eqref{bihamilgeneralflow}
which can be written down in a simple, explicit way in the case $\kappa=0$. 
We consider the algebraic scaling formulas \eqref{algHnegflow} and \eqref{algEnegflow},
where $q=1$, $s=0$, $\tilde s=6$ respectively are the scaling weights of $u$, $w$, $\tilde w$. 
For the first Hamiltonian density, 
we see $s-q+1=0$ implies that the algebraic scaling formula \eqref{algHnegflow} 
cannot be applied. 
However, since $D_xw=0$, 
we observe $H_{(1)} \equiv u w$ yields the Hamiltonian density
for the first Hamiltonian structure \eqref{bihamilgeneralflow}. 
For the second Hamiltonian density, 
the algebraic scaling formula is applicable since $\tilde s -q +1 \neq 0$. 
This yields 
$6E_{(1)} \equiv -D_x^{-1}(uD_x\tilde w)  = -D_x^{-1}(uh)  = -h_1$
after integration by parts, using equation \eqref{kuph1}. 
Hence $E_{(1)} \equiv -\tfrac{1}{6}h_1$ yields the Hamiltonian density
for the second Hamiltonian structure \eqref{bihamilgeneralflow}. 
Note these two Hamiltonian densities are similar in form to those for the mKdV $-1$ flow \eqref{mkdv-1flow}.

\subsection{KdV generalized -1 flow}

The recursion operator for the KdV hierarchy has the Hamiltonian factorization
\begin{equation}
\Rop=\Hop\Jop,
\quad
\Hop=D_x,
\quad
\Jop=D_x+\tfrac{1}{3}D_x^{-1}u+\tfrac{1}{3}uD_x^{-1}.
\end{equation}
We can write the generalized $-1$ flow equation \eqref{general-1flow} 
in the KdV hierarchy as a PDE system
\begin{subequations}\label{kdv-1flowsys}
\begin{gather}
 u_t=h,
\label{kdvh}\\
\Hop(w)=D_xw=\kappa h,
\label{kdvHw}\\
w=\Jop(h)=D_xh+\tfrac{1}{3}h_2+\tfrac{1}{3}uh_1,
\label{kdvJh}\\
D_xh_1=h,
\label{kdvh1}\\
D_xh_2=uh 
\label{kdvh2}
\end{gather}
\end{subequations}
involving $u$, a negative-flow variable $h$, and two potential variables $h_1,h_2$. 
To express this system equivalently in the form of a single non-evolutionary equation, 
we first substitute $w$ from equation \eqref{kdvJh} into equation \eqref{kdvHw},
\begin{equation}
D_x^2h+\tfrac{1}{3}u_xh_1+\tfrac{2}{3}uh-\kappa h=0. 
\end{equation}
Next, solving this equation for $h_1$, we obtain 
\begin{equation}\label{kdvh1expr}
h_1=u_x^{-1}((3\kappa-2u)h-3D_x^2h).
\end{equation}
Substituting this expression into equation \eqref{kdvh1}, 
and expanding out the terms, 
we get
\begin{equation}
-3D_x^3h+3u_x^{-1}u_{2x}D_x^2h+(3\kappa-2u)(D_xh-u_x^{-1}u_{2x}h)-3u_xh =0. 
\end{equation}
Finally, we substitute $h$ from equation \eqref{kdvh}, 
giving the equation 
\begin{equation}\label{kdv-1flow}
u_{t3x}=u_x^{-1}u_{2x}u_{t2x} +(\kappa-\tfrac{2}{3}u)(u_{tx}-u_x^{-1}u_{2x}u_t) -u_xu_t .
\end{equation}
This is the generalized KdV $-1$ flow. 

From \propref{firstintegral}, 
we can integrate the flow equation \eqref{kdv-1flow} 
once to obtain an equivalent equation of lower differential order, 
by use of the first integral \eqref{generalflowDxeqn}. 
The expression for the first integral is derived from the identities
\begin{equation}
w\Hop(w)=D_x\big(\tfrac{1}{2} w^2\big),
\quad
h\Jop(h)=h(D_xh+\tfrac{1}{3}h_2+\tfrac{1}{3}uh_1)=D_x\big(\tfrac{1}{2}h^2+\tfrac{1}{3}h_1h_2\big), 
\end{equation}
which together yield 
\begin{equation}\label{kdv-1flowid1}
D_x\big( \tfrac{1}{2} w^2 -\kappa  (\tfrac{1}{2}h^2+\tfrac{1}{3}h_1h_2) \big) =0
\end{equation}
holding for the PDE system \eqref{kdv-1flowsys}. 
The term involving $h_2$ in the identity  \eqref{kdv-1flowid1} 
can be eliminated as follows. 
We first combine equations \eqref{kdvHw} and \eqref{kdvh1}
to get $D_x(w-\kappa h_1)=0$. 
This relation yields
\begin{equation}\label{kdv-1flowid2}
D_x\big( \tfrac{1}{2} (w-\kappa h_1)^2 \big)=0 . 
\end{equation}
We next subtract the identities \eqref{kdv-1flowid2} and \eqref{kdv-1flowid1}, 
giving 
$D_x\big( \tfrac{1}{2}h^2 +\tfrac{1}{2} \kappa h_1^2 +h_1(\tfrac{1}{3}h_2 -w) \big) =0$,
and we then substitute equation \eqref{kdvJh}, 
which yields 
\begin{equation}
D_x\big( \tfrac{1}{2}h^2 +\tfrac{1}{2} \kappa h_1^2 -h_1(D_xh +\tfrac{1}{3}uh_1) \big) =0 . 
\end{equation}
This gives the first integral 
$h_1D_xh +(\tfrac{1}{3}u-\tfrac{1}{2} \kappa)h_1^2 -\tfrac{1}{2}h^2 =c(t)$
where $c(t)$ can be scaled by a conformal change of variable 
$t\rightarrow \tilde t(t)$ 
so that $c=\tilde{c}=\const$ without loss of generality. 
The resulting equation 
\begin{equation}
h_1D_xh +(\tfrac{1}{3}u-\tfrac{1}{2} \kappa)h_1^2 -\tfrac{1}{2}h^2 =\tilde{c}
\end{equation}
can be used to express 
\begin{equation}
h_1=\frac{-D_xh\pm\sqrt{(D_xh)^2+(\tfrac{2}{3}u-\kappa)(h^2+2\tilde{c})}}{\tfrac{2}{3}u-\kappa}.
\end{equation}
We now substitute this expression for $h_1$ along with $h=u_t$ into equation \eqref{kdvh1expr}, 
which yields
\begin{equation}\label{kdv-1flowintegrated} 
u_{t2x}=\frac{u_x}{2u-3\kappa}\Big( u_{tx}\mp\sqrt{u_{tx}^2+(\tfrac{2}{3}u-\kappa)(u_t^2+2\tilde{c})} \Big)-(\tfrac{2}{3}u-\kappa)u_t,
\end{equation}
which is equivalent to the KdV generalized $-1$ flow equation \eqref{kdv-1flow}
up to a conformal scaling of $t$. 

Finally, we state some main properties of this $-1$ flow. 

Firstly, 
the $-1$ flow equation \eqref{kdv-1flow} is invariant
under the scaling transformation 
$x\rightarrow\lambda x$, $t\rightarrow\lambda^{-3} t$, $u\rightarrow\lambda^{-2} u$, 
$\kappa\rightarrow\lambda^{-2}\kappa$. 
Secondly, 
from \thmref{symmsnegflow}, 
the $-1$ flow equation \eqref{kdv-1flow} possesses 
a hierarchy of symmetries \eqref{symmhierarchy}.
The root symmetry is an $x$-translation $\X^{(0)}=u_x\p_u$,
and the first higher-order symmetry 
\begin{equation}
\X^{(1)}=(u_{3x} +uu_x)\p_u
\end{equation}
corresponds to the KdV flow $u_t= u_{3x} +uu_x$. 
Thirdly, 
from \thmref{conslawsnegflow}, 
the $-1$ flow equation \eqref{kdv-1flow} also possesses 
a hierarchy of global conservation laws \eqref{conslawhierarchy} 
which are related to the well-known conservation laws of the KdV flow. 
Recall, the first four KdV conservation laws are given by 
the Hamiltonian (conserved) densities 
$H^{(0)}=u$, 
$H^{(1)}=\tfrac{1}{2}u^2$, 
$H^{(2)}=-\tfrac{1}{2}u_x^2 + \tfrac{1}{6}u^3$, 
and $H^{(3)}=\tfrac{1}{2}u_{2x}^2 -\tfrac{5}{6}uu_x^2 + \tfrac{5}{72}u^4$. 
This yields the corresponding conserved densities
\begin{gather}
\tilde{H}^{(0)}=\tfrac{1}{2} u^2 -\kappa u
\\
\tilde{H}^{(1)}=-\tfrac{1}{2}( u_x^2 -\tfrac{1}{3}u^3 +\kappa u^2 )
\\
\tilde{H}^{(2)}=\tfrac{1}{2}( u_{2x}^2 -\tfrac{5}{3}uu_x^2 + \tfrac{5}{36}u^4 +\kappa (u_x^2 -\tfrac{1}{3}u^3) )
\end{gather}
for the $-1$ flow equation \eqref{kdv-1flow}. 
Fourthly, 
from \thmref{biHamilnegflow}, 
the $-1$ flow equation \eqref{kdv-1flow} has a bi-Hamiltonian structure \eqref{bihamilgeneralflow}. 
This structure can be written down in a simple, explicit way in the case $\kappa=0$
by using the algebraic scaling formulas \eqref{algHnegflow} and \eqref{algEnegflow}. 
Note $q=2$, $s=0$, $\tilde s=2$ respectively are the scaling weights of $u$, $w$, $\tilde w$. 
Hence the first Hamiltonian density is given by 
$-H_{(1)} \equiv -2u w +D_x^{-1}(u_x w) \equiv -u w$
after integration by parts, using $D_xw=0$. 
This yields $H_{(1)} \equiv u w$. 
For the second Hamiltonian density, 
we obtain 
$E_{(1)} \equiv -2u\tilde w +D_x^{-1}(u_x\tilde w)  \equiv -u\tilde w -h_2$
after integration by parts, using equation \eqref{kdvh2}. 

We remark that when $\kappa=0$ the generalized KdV $-1$ flow equation \eqref{kdv-1flowintegrated} 
reduces to the ordinary mKdV $-1$ flow 
which has been studied in the special case $\tilde{c}=0$ in \Ref{QiaLi}. 
This equation 
$u_{t2x}=\tfrac{1}{2}u^{-1}u_x\big( u_{tx}\mp\sqrt{u_{tx}^2+\tfrac{2}{3}u u_t^2} \big)-\tfrac{2}{3}u u_t$
can be written in the simpler form 
$v_{t2x}= v^{-1}v_tv_{2x}+2 v^2 v_x$ 
through the use of the transformation $u=-6v_{2x}/v$ \cite{QiaLi}, 
up to a scaling of $v$.

\subsection{Sawada-Kotera generalized $-1$ flow}

The recursion operator for the Sawada-Kotera hierarchy has the Hamiltonian factorization $\Rop=\Hop\Jop$ given by 
\begin{align}
&\Hop=D_x^3+2D_x(D_x^{-1}u+uD^{-1}_x)D_x=D_x^3+4uD_x+2u_x,
\\
&\Jop=(D_x+D_x^{-1}u)D_x(D_x+uD_x^{-1})=D_x^3 +2uD_x +u_x +(\tfrac{1}{2}u^2+u_{2x})D_x^{-1} + D_x^{-1}(\tfrac{1}{2}u^2+u_{2x}) . 
\end{align}
Note 
\begin{equation}
\Jop(h)=\A+\B D_x^{-1}h+D_x^{-1}(\B h), 
\end{equation}
where
\begin{equation}
\A=D_x^3h+2uD_xh+u_xh,
\quad
\B=u_{2x}+\tfrac{1}{2}u^2 . 
\end{equation}
Then we can write the generalized $-1$ flow equation \eqref{general-1flow} 
in the Sawada-Kotera hierarchy as a PDE system
\begin{subequations}\label{sk-1flowsys}
\begin{gather}
u_t=h,
\label{skh}\\
\Hop(w)=D_x^3 w +4uD_xw+2u_xw =\kappa h,
\label{skHw}\\
w=\Jop(h)= \A+\B h_1+h_2,
\label{skJh}\\
D_xh_1=h,
\label{skh1}\\
D_xh_2=\B h
\label{skh2}
\end{gather}
\end{subequations}
which involves $u$, a negative-flow variable $h$, and two potential variables $h_1,h_2$. 
This system is equivalent to a single non-evolutionary equation 
which is obtained by the following steps. 

First, we substitute $w$ from equation \eqref{skJh} into equation \eqref{skHw}
and simplify the terms by using 
\begin{equation}\label{skDw}
D_x w=\C_1+h_1D_x\B,
\quad
D_x^2w=\C_2+h_1D_x^2\B,
\quad
D_x^3w=\C_3+h_1D_x^3\B
\end{equation}
where
\begin{gather}
\C_1=D_x\A+2\B h,
\quad
\C_2=D_x^2\A+3(D_x\B)h+2\B D_xh,
\label{skC1C2}\\
\C_3=D_x^3\A+4(D_x^2\B)h+5(D_x\B)D_xh+2\B D_x^2h.
\label{skC1C3}
\end{gather}
This yields
\begin{equation}
\C_3+4u\C_1+2u_x\A +(D_x^3\B+4uD_x\B+2u_x\B)h_1 +2u_x h_2 = \kappa h . 
\end{equation}
Next, we solve this equation for $h_2$, which gives 
\begin{equation}\label{skh2expr}
h_2=-\frac{\F h_1 +\G -\kappa h}{2u_x}
\end{equation}
where
\begin{equation}
\F=D_x^3\B+4uD_x\B+2u_x\B,
\quad
\G=\C_3+4u\C_1+2u_x\A . 
\end{equation}
We substitute this expression \eqref{skh2expr} into equation \eqref{skh2}, 
yielding
\begin{equation}\label{skh1eqn}
\I h_1 +u_{2x}G-u_xD_xG -\K -\kappa\J =0
\end{equation}
where 
\begin{equation}
\I = u_{2x}F-u_xD_x F,
\quad
\J=u_{2x}h -u_xD_x h,
\quad
\K=u_x(2u_xB+F)h . 
\end{equation}
Then, we solve equation \eqref{skh1eqn} for 
\begin{align}\label{skh1expr}
h_1=\frac{u_xD_x\G -u_{2x}\G +\K +\kappa\J}{\I}
\end{align}
and substitute it into equation \eqref{skh1}. 
This gives 
\begin{equation}\label{skeqn}
-D_x\I (u_x D_x\G -u_{2x}\G+\K +\kappa\J)+\I (u_x D_x^2\G - u_{3x}\G +D_x\K +\kappa D_x\J ) -\I^2 h =0 . 
\end{equation}
The highest derivative term in equation \eqref{skeqn} 
comes from the $D_x^2\G$ term,
with the highest derivative term in $\G$ coming from $D_x^3A$ in $\C_3$,
where the highest derivative term in $D_x^3A$ is given by $D_x^6h$. 
If we isolate the highest derivative term in $G$ by writing
\begin{equation}\label{skhighestder}
\G=D_x^6h+\M
\end{equation}
where 
\begin{equation}
\M = 4u\C_1+2u_x\A + D_x^3(u_xh+2uD_xh) +4(D_x^2\B)h+5(D_x\B)D_xh+2\B D_x^2h, 
\end{equation}
then the highest derivative term in $D_x^2G$ is $D_x^8h$.
Isolating this term in equation \eqref{skeqn} 
and substituting $h$ from equation \eqref{skh},
we get the generalized SK $-1$ flow equation
\begin{equation}\label{sk-1flow}
\begin{aligned}
u_{t8x}=&
\frac{D_x\I (u_{t7x}+D_x\M)}{\I} 
-\frac{D_x\I (u_{2x}(u_{t6x}+\M)-\K -\kappa\J)}{u_x\I}
\\&\qquad
+\frac{u_{3x}(u_{t6x}+\M) -D_x\K -\kappa D_x\J +\I u_t}{u_x} -D_x^2\M 
\end{aligned}
\end{equation}
with 
\begin{align}
&\begin{aligned}
\M & = 6u u_{t4x} +9u_x u_{t3x} +(9u^2+11u_{2x})u_{t2x} +(21uu_x+10u_{3x})u_{tx} 
\\&\qquad
+(4u^3+6u_x^2 +16 uu_{2x}+5u_{4x})u_t , 
\end{aligned}
\label{skM}\\
& \I = -u_x u_{6x} +u_{2x} u_{5x} -5 u u_x u_{4x} +5 (u u_{2x} -2 u_x^2) u_{3x} -10 u u_x^3 , 
\label{skI}\\
& \K = u_x(u_{5x}+5u u_{3x} +7u_x u_{2x}+6 u^2 u_x)u_t, 
\label{skK}\\
& \J= u_{2x}u_t - u_x u_{tx} . 
\label{skJ}
\end{align}

From \propref{firstintegral}, 
we can use the first integral \eqref{generalflowDxeqn}
to integrate the flow equation \eqref{sk-1flow} 
once to obtain an equivalent equation of lower differential order. 
The expression for the first integral is derived from the identities
\begin{align}
&
w\Hop(w)=wD_x^3w+4uwD_xw+2u_xw^2=D_x(wD_x^2w-\tfrac{1}{2}(D_xw)^2+2uw^2),
\\
&
hw=hD_x^3h+2uhD_xh+u_xh^2+\B hh_1+hh_2=D_x(hD_x^2h-\tfrac{1}{2}(D_xh)^2+uh^2+h_1h_2).
\end{align}
Combining these identities, we get
\begin{equation}
D_x\big( wD_x^2w-\tfrac{1}{2}(D_xw)^2+2uw^2-\kappa(hD_x^2h-\tfrac{1}{2}(D_xh)^2+uh^2+h_1h_2) \big)=0,
\end{equation}
which yields the first integral 
\begin{equation}\label{sk1stintegral}
\begin{aligned}
& (\A+\B h_1+h_2)((D_x^2\B)h_1 +\C_2)
-\tfrac{1}{2}((D_x\B)h_1 +\C_1)^2 
\\&\qquad
+2u(\A+\B h_1+h_2)^2
-\kappa(hD_x^2h-\tfrac{1}{2}(D_xh)^2+uh^2+h_1h_2)=c(t)
\end{aligned}
\end{equation}
holding for the PDE system \eqref{sk-1flowsys}. 
By a conformal change of variable $t\rightarrow\tilde{t}$, 
we can put $c=\tilde{c}=\const$ without loss of generality. 
Then, we substitute $h_2$ from equation \eqref{skh2expr} 
into this first integral \eqref{sk1stintegral}, 
which gives a quadratic equation in $h_1$, 
\begin{equation}\label{skh1quadreqn}
\N h_1^2 + \P h_1 + \R =0
\end{equation}
where
\begin{align}
& \N = u(2u_x\B-\F)^2 +u_x( (2u_x\B-\F)D_x^2\B -u_x(D_x\B)^2 +\kappa\F), 
\\
&\begin{aligned}
\R & = u(2u_x\A -\G+\kappa h)^2 +u_x^2(2\tilde c +\C_1^2)
+u_x\C_2(2u_x\A -\G+\kappa h) 
\\&\qquad
-\kappa u_x^2(2(D_x^2 h+uh)h -(D_xh)^2) , 
\end{aligned}
\\
&\begin{aligned}
\P & = (2u_x\A -\G+\kappa h)(u_xD_x^2\B+2u(2u_x\B-\F)) +u_x(2u_x\B-\F)\C_2
\\&\qquad
-2u_x^2(D_x\B)\C_1
+\kappa u_x(\G-\kappa h) . 
\end{aligned}
\end{align}
We solve equation \eqref{skh1quadreqn}
and substitute $h_1$ into equation \eqref{skh1expr}. 
This yields
\begin{equation}\label{skeqnlowerord}
2(u_xD_x\G -u_{2x}\G +\K +\kappa\J)\N=(-\P\pm\sqrt{\P^2-4\R\N})\I . 
\end{equation}
The highest derivative term in this equation comes from $D_x\G$,
which gives, using equation \eqref{skhighestder}, 
\begin{equation}
D_x\G=D_x^7h+D_x\M . 
\end{equation}
Hence, 
by solving equation \eqref{skeqnlowerord} for $D_x^7h$ and then substituting $h=u_t$, 
we obtain the lower-order non-evolutionary equation 
\begin{equation}\label{sk-1flowintegrated}
u_{t7x}=
\frac{u_{2x}(u_{t6x}+\M) -\K-\kappa\J}{u_x} 
-\frac{(\P\mp\sqrt{\P^2-4\R\N})\I}{2u_x\N} 
-D_x\M
\end{equation}
where
\begin{align}
& \N=  u\N_1^2 -u_x\N_2 +\kappa u_x \F , 
\\
& \P= u_x\C_2\N_1 +\P_1\P_2-2u_x^2(u_{3x}+uu_x)\C_1 +\kappa u_x\P_3, 
\\
&\R = u\P_1^2-u_x\C_2\N_1 -u_x^2(\C_1^2+2\tilde{c}) -\kappa u_x^2(2u_t(u_{t2x}+uu_t)-u_{tx}^2), 
\\
& \F = u_{5x} +5uu_{3x}+5u_xu_{2x}+5u^2u_x, 
\\
& \C_1= u_{t4x} +2uu_{t2x} +3u_xu_{tx} +(3u_{2x}+u^2)u_t, 
\\
& \C_2 = u_{t5x} +2uu_{t3x} +5u_xu_{t2x} +(6u_{2x}+u^2)u_{tx} +(4u_{3x}+3uu_x)u_t, 
\\
& \N_1=u_{5x} +5uu_{3x}+3u_xu_{2x}+4u^2u_x, 
\\
&\begin{aligned}
\N_2 &= 
(u_{4x}+uu_{2x}+u_x^2) u_{5x} +(5uu_{3x}+3u_xu_{2x}+4u^2u_x) u_{4x}
+u_xu_{3x}^2 +(5u^2u_{2x}+7uu_x^2)u_{3x} 
\\&\qquad
+3uu_xu_{2x}^2 +(3u_x^2+4u^3)u_xu_{2x} +5u^2u_x^3 , 
\end{aligned}
\\
&\begin{aligned}
\P_1 & = 
u_{t6x}+6uu_{t4x}+7u_xu_{t3x}+(11u_{2x}+9u^2)u_{t2x}+(10u_{3x}+17uu_x)u_{tx}
\\
&\qquad
+(5u_{4x}+16uu_{2x}+4u_x^2+4u^3-\kappa)u_t, 
\end{aligned}
\\
& \P_2 = 2uu_{5x} -u_xu_{4x} +10u^2u_{3x} +5uu_xu_{2x}+8u^3u_x -u_x^3, 
\\
&\begin{aligned}
\P_3 & = 
u_{t6x}+6uu_{t4x}+9u_xu_{t3x}+(11u_{2x}+9u^2)u_{t2x}+(10u_{3x}+21uu_x)u_{tx}
\\
&\qquad
+(5u_{4x}+16uu_{2x}+6u_x^2+4u^3-\kappa)u_t, 
\end{aligned}
\end{align}
and where $\M$, $\I$, $\J$, $\K$ are given by expressions \eqref{skM}--\eqref{skK}.
This equation \eqref{sk-1flowintegrated} is equivalent to the generalized SK $-1$ flow equation \eqref{sk-1flow} 
up to a conformal scaling of $t$. 

Finally, we state some main properties of this $-1$ flow. 

Firstly, 
the $-1$ flow equation \eqref{sk-1flow} is invariant
under the scaling transformation 
$x\rightarrow\lambda x$, $t\rightarrow\lambda^{-5} t$, $u\rightarrow\lambda^{-2} u$, 
$\kappa\rightarrow\lambda^{-2}\kappa$. 
Secondly, 
from \thmref{symmsnegflow}, 
the $-1$ flow equation \eqref{sk-1flow} possesses 
a hierarchy of symmetries \eqref{symmhierarchy}.
The root symmetry is an $x$-translation $\X^{(0)}=u_x\p_u$,
and the first higher-order symmetry 
\begin{equation}
\X^{(1)}=(u_{5x}+5uu_{3x}+5u_{x}u_{2x}+5u^2u_{x})\p_u
\end{equation}
corresponds to the SK flow 
$u_t=u_{5x}+5uu_{3x}+5u_{x}u_{2x}+5u^2u_{x}$. 
Thirdly, 
from \thmref{conslawsnegflow}, 
the $-1$ flow equation \eqref{sk-1flow} also possesses 
a hierarchy of global conservation laws \eqref{conslawhierarchy} 
which are related to the conservation laws of the SK flow. 
The first three SK conservation laws \cite{FucOev}
are given by the Hamiltonian (conserved) densities 
$H^{(0)}=u$, 
$H^{(1)}=-\tfrac{1}{2}u_x^2 + \tfrac{1}{6}u^3$, 
and $H^{(2)}=\tfrac{1}{2}u_{2x}^2 -\tfrac{3}{2}uu_x^2 + \tfrac{1}{6}u^4$. 
This yields the corresponding conserved densities
\begin{gather}
\tilde{H}^{(0)}=-\tfrac{1}{2}( u_x^2 -\tfrac{1}{3}u^3 +\kappa u )
\\
\tilde{H}^{(1)}=\tfrac{1}{2}( u_{2x}^2 -3uu_x^2 + \tfrac{1}{3}u^4 +\kappa (u_x^2 -\tfrac{1}{3}u^3) )
\end{gather}
for the $-1$ flow equation \eqref{sk-1flow}. 
Fourthly, 
from \thmref{biHamilnegflow}, 
the $-1$ flow equation \eqref{sk-1flow} has a bi-Hamiltonian structure \eqref{bihamilgeneralflow}. 
Note $q=2$, $s=\tilde s=1$ respectively are the scaling weights of $u$, $w$, $\tilde w$.
Since we have $s-q+1=\tilde s-q+1=0$, 
the algebraic scaling formulas \eqref{algHnegflow} and \eqref{algEnegflow}
cannot be applied here.

\subsection{Kaup-Kupershmidt generalized $-1$ flow}
The recursion operator for the Kaup-Kupershmidt hierarchy has the Hamiltonian factorization $\Rop=\Hop\Jop$,
where
\begin{align}
& \Hop = D_x^3+\tfrac{1}{2}D_x(D_x^{-1}u+uD_x^{-1})D_x
=D_x^3+uD_x+\tfrac{1}{2}u_x,
\\
&\begin{aligned}
\Jop & =D_x^3+\tfrac{3}{2}(uD_x+D_xu)+D_x^2uD_x^{-1}+D_x^{-1}uD_x^2+2(u^2D_x^{-1}+D_x^{-1}u^2)
\\
& =D_x^3+5uD_x+\tfrac{5}{2}u_x+(u_{2x}+2u^2)D_x^{-1}+D_x^{-1}(u_{2x}+2u^2).
\end{aligned}
\end{align}
Note
\begin{equation}
\Jop(h)=\A+\B D_x^{-1}h+D_x^{-1}(\B h),
\end{equation}
where
\begin{align}
\A=D_x^3h+5uD_xh+\tfrac{5}{2}u_xh,
\quad
\B=u_{2x}+2u^2.
\end{align}
We can write the generalized $-1$ flow equation \eqref{general-1flow} in the Kaup-Kupershmidt hierarchy as the following PDE system
\begin{subequations}\label{kk-1flowsys}
\begin{gather}
u_t=h,
\label{kkh}\\
\Rop(h)=\Hop(w)=D_x^3w+uD_xw+\tfrac{1}{2}u_xw=\kappa h,
\label{kkHw}\\
w=\Jop(h)=\A+\B h_1+h_2,
\label{kkJh}\\
D_xh_1=h,
\label{kkh1}\\
D_xh_2=\B h
\label{kkh2}
\end{gather}
\end{subequations}
involving $u$, a negative-flow variable $h$, and two potential variables $h_1,h_2$. 
This system is very similar to the system \eqref{sk-1flowsys} for the generalized SK $-1$ flow. 

We now write the system \eqref{kk-1flowsys} as single non-evolutionary equation 
by the following steps.
First, we substitute $w$ from equation \eqref{kkJh} into equation \eqref{kkHw} and simplify the terms by using equations \eqref{skDw}--\eqref{skC1C2} to get 
\begin{equation}
\C_3+u\C_1+\tfrac{1}{2}u_x\A+(D_x^3\B+uD_x\B+\tfrac{1}{2}u_x\B)h_1+\tfrac{1}{2}u_xh_2=\kappa h,
\end{equation}
where
\begin{align}
\G=\C_3+u\C_1+\tfrac{1}{2}u_x\A,
\quad
\F=D_x^3B+uD_xB+\tfrac{1}{2}u_x\B. 
\end{align}
Next, we solve this equation for $h_2$, which gives
\begin{equation}\label{kkh2expr}
h_2=\frac{2(\kappa h-\G-\F h_1)}{u_x}.
\end{equation}
We substitute this expression \eqref{kkh2expr} into the equation \eqref{kkh2}, yielding
\begin{equation}\label{kkh1eqn}
\I h_1+u_{2x}\G-u_xD_x\G-\K-\kappa\J=0,
\end{equation}
where
\begin{gather}
\I=u_{2x}\F-u_xD_x\F,
\quad
\K=(u_x\F+\tfrac{1}{2}u_x^2\B)h,
\quad
\J=u_{2x}h-u_xD_xh.
\end{gather}
Then, we solve \eqref{kkh1eqn} for
\begin{equation}\label{kkh1exp}
h_1=\frac{u_xD_x\G-u_{2x}\G+\K+\kappa\J}{\I}
\end{equation}
and substitute it into equation \eqref{kkh1}.
This gives
\begin{equation}\label{kkeqn}
-D_x\I(u_xD_x\G-u_{2x}\G+\K+\kappa\J)+\I(u_xD_x^2\G-u_{3x}\G+D_x\K+\kappa D_x\J)-\I^2h=0 . 
\end{equation}
The highest derivative term in equation \eqref{kkeqn} comes from the $D_x^2\G$ term,
with the highest derivative term in $\G$ coming from $D_x^3\A$ in $\C_3$,
where the highest term in $D_x^3\A$ is given by $D_x^6h$.
If we isolate the highest derivative term in $\G$ by writing
\begin{equation}
\G=D_x^6h+\M,
\end{equation}
where
\begin{align}
\M=u\C_1+\tfrac{1}{2} u_x\A+5D_x^3(uD_xh+\tfrac{1}{2}u_xh)+4(D_x^2\B)h+5(D_x\B)D_xh+2\B D_x^2h,
\end{align}
then the highest derivative term in $D_x^2\G$ will be $D_x^8h$.
Isolating this term in equation \eqref{kkeqn} and substituting $h$ from equation \eqref{kkh},
we get the generalized KK $-1$ flow equation
\begin{equation}\label{kk-1flow}
\begin{aligned}
u_{t8x}=&
\frac{D_x\I(u_{t7x}+D_x\M)}{\I}-\frac{D_x\I(u_{2x}(u_{t6x}+\M)-\K-\kappa\J}{u_x\I}
\\&\qquad
+\frac{u_{3x}(u_{t6x}+\M)-D_x\K-\kappa D_x\J+\I u_t}{u_x}-D_x^2\M,
\end{aligned}
\end{equation}
where
\begin{align}
&\begin{aligned}
\M &=
6uu_{t4x}+18u_xu_{t3x}+(9u^2+\tfrac{49}{2}u_{2x})u_{t2x}
+5(\tfrac{7}{2}u_{3x}+6uu_x)u_{tx}
\\&\qquad
+(\tfrac{13}{2}u_{4x}+\tfrac{41}{2}uu_{2x}+\tfrac{69}{4}u_x^2+4u^3)u_t, 
\end{aligned}
\label{kkM}\\
& \I=-u_xu_{6x}+u_{2x}u_{5x}-5uu_xu_{4x}+5(uu_{2x}-\tfrac{7}{2}u_x^2)u_{3x}-10uu_x^3 , 
\label{kkI}\\
& \K=u_x(u_{5x}+5uu_{3x}+13u_xu_{2x}+6u^2u_x)u_t, 
\label{kkK}\\
& \J=u_{2x}u_t-u_xu_{tx} . 
\label{kkJ}
\end{align}

From \propref{firstintegral},
we can use the first integral \eqref{generalflowDxeqn}
to integrate the flow equation \eqref{kk-1flow}
once to obtain an equivalent equation of lower differential order.
The expression for the first integral is derived from the identities
\begin{align}
& w\Hop(w) = w(D_x^3w+uD_xw+\tfrac{1}{2}u_xw)
=D_x(w D_x^2w-\tfrac{1}{2}(D_xw)^2+\tfrac{1}{2}w^2u),
\\
&\begin{aligned}
hw & = h(D_x^3h+5uD_xh+\tfrac{5}{2}u_xh+(u_{2x}+2u^2)h_1+h_2)\\
& =D_x(hD_x^2h-\tfrac{1}{2}(D_xh)^2+\tfrac{5}{2}uh^2+h_1h_2) , 
\end{aligned}
\end{align}
which combine to give
\begin{equation}
D_x(w D_x^2w-\tfrac{1}{2}(D_xw)^2+\tfrac{1}{2}w^2u-\kappa(hD_x^2h-\tfrac{1}{2}(D_xh)^2+\tfrac{5}{2}uh^2+h_1h_2))=0 . 
\end{equation}
This yields the first integral
\begin{equation}\label{kk1stintegral}
\begin{aligned}
&(\A+\B h_1+h_2)(\C_2+(D_x^2\B)h_1)-\tfrac{1}{2}(\C_1+(D_x\B)h_1)^2+\tfrac{1}{2}u(\A+\B h_1+h_2)^2
\\
&\qquad
-\kappa(hD_x^2h-\tfrac{1}{2}(D_xh)^2+\tfrac{5}{2}uh^2+h_1h_2)=c(t)
\end{aligned}
\end{equation}
holding for the PDE system \eqref{kk-1flowsys}.
By a conformal change of variable $t\rightarrow\tilde{t}$,
we can put $c=\tilde{c}=\const$ without loss of generality.
Then, we substitute $h_2$ from equation \eqref{kkh2expr}
into this first integral \eqref{kk1stintegral},
which gives a quadratic equation in $h_1$,
\begin{equation}\label{kkh1quadreqn}
\N h_1^2 + \P h_1 + \R =0,
\end{equation}
where
\begin{align}
& \N=u(\B u_x-2\F)^2+2u_x(D_x^2\B)(\B u_x-2\F)-u_x^2(D_x\B)^2+4\kappa u_x\F . 
\\
&\begin{aligned}
\P=&2(\A u_x+2\kappa h-2\G)(u_xD_x^2\B+u(\B u_x-2\F))-2u_x^2(D_x\B)\C_1
\\&\qquad
+2u_x\C_2(\B u_x-2\F) +4\kappa(\G-\kappa h)u_x , 
\end{aligned}
\\
&\begin{aligned}
\R=&u(\A u_x+2\kappa h-2\G)^2+2u_x(\A u_x+2\kappa h-2\G)\C_2-u_x^2\C_1^2
\\&\qquad
-u_x^2(\kappa(2hD_x^2h-(D_xh)^2+5uh^2)+2\tilde{c}) . 
\end{aligned}
\end{align}
We solve the equation \eqref{kkh1quadreqn} for $h_1$ and substitute it into the equation \eqref{kkh1exp},
yielding 
\begin{equation}\label{kkeqnlowerord}
2(u_xD_x\G-u_{2x}\G+\K+\kappa\J)\N=(-\P\pm\sqrt{\P^2-4\N\R})\I . 
\end{equation}
The highest derivative term in this equation comes from $D_x\G$, which gives 
\begin{align}
D_x\G=D_x^7h+D_x\M . 
\end{align}
Hence, by solving equation \eqref{kkeqnlowerord} for $D_x^7h$ and then substituting $h=u_t$, 
we obtain the lower-order non-evolutionary equation
\begin{equation}\label{kk-1flowintegrated}
u_{t7x}=\frac{u_{2x}(u_{t6x}+\M)-\K-\kappa\J}{u_x}-\frac{(\P\mp\sqrt{\P^2-4\N\R})I}{2u_x\N}-D_x\M
\end{equation}
where 
\begin{align}
& \N=u\N_1^2+u_x\N_2+4\kappa u_x\F,
\\
& \R=u\R_1^2+2u_x\R_1\C_2-u_x^2\C_1^2-u_x^2(\kappa(2u_tu_{t2x}-u_{tx}^2+5uu_t^2)+2\tilde{c}), 
\\
& \P=2\R_1\P_1+2u_x\N_1\C_2-2u_x^2(u_{3x}+4uu_x)\C_1 +4\kappa\P_2, 
\\
&\F=u_{5x}+5uu_{3x}+\tfrac{25}{2}u_xu_{2x}+5u^2u_x, 
\\
&\C_1=u_{t4x}+5uu_{t2x}+\tfrac{15}{2}u_xu_{tx}+(4u^2+\tfrac92u_{2x})u_t, 
\\
&\C_2=u_{t5x}+5uu_{t3x}+\tfrac{25}{2}u_xu_{t2x}+4(3u_{2x}+u^2)u_{tx}+(\tfrac{11}{2}u_{3x}+12uu_x)u_t, 
\\
&\N_1 =-2(u_{5x}+5uu_{3x}+12u_xu_{2x}+8u^2u_x), 
%=\B u_x-2\F
\\
&\begin{aligned}
\N_2 =&
-4(u_{4x}+4uu_{2x}+4u_x^2)u_{5x}-4(5uu_{3x}+12u_xu_{2x}+4u^2u_x)u_{4x}-u_xu_{3x}^2
\\&\qquad
-8u(11u_x^2+10uu_{2x})u_{3x}-192uu_xu_{2x}^2-64(3u_x^2+u^3)u_xu_{2x}-80u^2u_x^3, 
%=2(D_x^2\B)\N_1-u_x(D_x\B)^2
\end{aligned}
\\
&\begin{aligned}
\R_1=&
-2u_{t6x}-12uu_{t4x}-35u_xu_{t3x}-(18u^2+49u_{2x})u_{t2x}-5(7u_{3x}+11uu_x)u_{tx}
\\&\qquad
-(13u_{4x}+41uu_{2x}+32u_x^2+8u^3-2\kappa)u_t, 
%=\A u_x+2\kappa h-2\G
\end{aligned}
\\
& \P_1 =-2uu_{5x}+u_xu_{4x}-10u^2u_{3x}-20uu_xu_{2x}+4u_x^3-8u^3u_x, 
%=u_xD_x^2B+u\N_1
\\
&\begin{aligned}
\P_2=& 
u_{t6x} + 6uu_{t4x}+18u_xu_{t3x}+(9u^2+\tfrac{49}{2}u_{2x})u_{t2x}
+5(\tfrac{7}{2}u_{3x}+6uu_x)u_{tx}
\\&\qquad
+(\tfrac{13}{2}u_{4x}+\tfrac{41}{2}uu_{2x}+\tfrac{69}{4}u_x^2+4u^3-\kappa )u_t, 
\end{aligned}
\end{align}
and where $\M$, $\I$, $\J$, $\K$ are given by expressions \eqref{kkM}--\eqref{kkK}.
This equation \eqref{kk-1flowintegrated} is equivalent to the generalized KK $-1$ flow equation \eqref{kk-1flow} 
up to a conformal scaling of $t$. 

Finally, we state some main properties of this $-1$ flow. 

Firstly, 
the $-1$ flow equation \eqref{kk-1flow} is invariant
under the scaling transformation 
$x\rightarrow\lambda x$, $t\rightarrow\lambda^{-5} t$, $u\rightarrow\lambda^{-2} u$, 
$\kappa\rightarrow\lambda^{-2}\kappa$. 
Secondly, 
from \thmref{symmsnegflow}, 
the $-1$ flow equation \eqref{kk-1flow} possesses 
a hierarchy of symmetries \eqref{symmhierarchy}.
The root symmetry is an $x$-translation $\X^{(0)}=u_x\p_u$,
and the first higher-order symmetry 
\begin{equation}
\X^{(1)}=(u_{5x}+5uu_{3x}+\tfrac{25}{2}u_{x}u_{2x}+5u^2u_{x})\p_u
\end{equation}
corresponds to the KK flow 
$u_t=u_{5x}+5uu_{3x}+\tfrac{25}{2}u_{x}u_{2x}+5u^2u_{x}$. 
Thirdly, 
from \thmref{conslawsnegflow}, 
the $-1$ flow equation \eqref{kk-1flow} also possesses 
a hierarchy of global conservation laws \eqref{conslawhierarchy} 
which are related to the conservation laws of the KK flow. 
The first three KK conservation laws \cite{SatKau}
are given by the Hamiltonian (conserved) densities 
$H^{(0)}=u$, 
$H^{(1)}=-\tfrac{1}{2}u_x^2 +\tfrac{2}{3}u^3$, 
and $H^{(2)}=\tfrac{1}{2}u_{2x}^2 -3uu_x^2+ \tfrac{2}{3}u^4$.
This yields the corresponding conserved densities
\begin{gather}
\tilde{H}^{(0)}=-\tfrac{1}{2}( u_x^2 -\tfrac{2}{3}u^3 +\kappa u )
\\
\tilde{H}^{(1)}=\tfrac{1}{2}( u_{2x}^2 -6uu_x^2 + \tfrac{4}{3}u^4 +\kappa (u_x^2 -\tfrac{2}{3}u^3) )
\end{gather}
for the $-1$ flow equation \eqref{kk-1flow}. 
Fourthly, 
from \thmref{biHamilnegflow}, 
the $-1$ flow equation \eqref{sk-1flow} has a bi-Hamiltonian structure \eqref{bihamilgeneralflow}. 
Note $q=2$, $s=\tilde s=1$ respectively are the scaling weights of $u$, $w$, $\tilde w$.
Since we have $s-q+1=\tilde s-q+1=0$, 
the algebraic scaling formulas \eqref{algHnegflow} and \eqref{algEnegflow}
cannot be applied here,
similarly to the SK $-1$ flow.

\section{Concluding remarks}
\label{remarks}

We have introduced a one-parameter generalization \eqref{generalnegflowhierarchy} of
the hierarchy of negative flows \eqref{negflowhierarchy} associated with each hierarchy of 
integrable scalar evolution equations \eqref{integrhierarchy} of semilinear polynomial form. 
As main results, 
several important properties of this generalization have been established, 
which also apply to the hierarchy of ordinary negative flows,
and for each hierarchy the first generalized negative flow has been worked out explicitly. 

These generalized negative flows provide a wider (new) class of 
non-evolutionary integrable nonlinear wave equations. 
Other negative flow hierarchies, 
such as for the Boussinesq equation and the Landau-Lifshitz equations, 
can be generalized in a similar way. 

There are some interesting directions in which our work in the present paper 
can be extended. 

First, 
the (multi) soliton solutions of the generalized negative flow equations in each hierarchy 
can be derived and their interaction properties can be studied and compared to 
the ordinary negative flow solitons as well as to the solitons of the root equation. 
Second, 
a Lax pair can be sought for each hierarchy of generalized negative flow equations,
using the methods in \Ref{Qia}. 
Third, 
when the root equation in a hierarchy has an associated ``dual-type'' peakon equation,
which arises from splitting the Hamiltonian operators of the root equation \cite{OlvRos},
a relation between the peakon equation and the first generalized negative flow equation 
can be explored by looking for a hodograph transformation, 
as occurs in the KdV case \cite{Fuc96}. 

Finally, 
it could be interesting to study a larger generalization of the integrable hierarchies 
by combining together both negative and positive flows into a single equation
$\Rop^n(u_t)=\kappa_n u_t + \tilde\kappa_m \Rop^m(u_x)$, 
$n=1,2,\ldots$ and $m=1,2,\ldots$, 
with $\kappa_n$ and $\tilde\kappa_m$ being (constant) parameters.

\section*{Acknowledgements}

S.C.A. and T.W. are each supported by an NSERC Discovery grant. 
C.Z. is supported by National Natural Science Foundation of  China (Grant No.\ 11301007) 
and the Natural Science Foundation of Anhui province (Grant No.\ 1408085QA05). 

T. Tsuchida and Z. Qiao are thanked for helpful remarks.


\begin{thebibliography}{99}

\bibitem{Fok} 
A.S. Fokas, 
A symmetry approach to exactly solvable evolution equations, 
{\em J. Math. Phys.} {\bf 21} (1980), 1318--1325. 

\bibitem{IbrSha} 
N. Ibragimov and A.B. Shabat, 
Evolutionary equations with nontrivial Lie-Backlund algebra, 
{\em Funct. Anal. Appl.} {\bf 14} (1980), 19--30.

\bibitem{Mag}
F. Magri,
A simple model of the integrable Hamiltonian equation,
{\em J. Math. Phys.} {\bf 19} (1978), 1156--1162. 

\bibitem{Olv}
P.J. Olver, 
{\em Applications of the groups to differential equations} 
(2nd ed.) Springer-Verlag, 1993.

\bibitem{Lax} 
P.D. Lax,
Integrals of nonlinear equations of evolution and solitary waves, 
{\em Commun. Pure Appl. Math.} {\bf 21} (1968), 467--490. 

\bibitem{ZakSha}
V.E. Zakharov and A.B. Shabat. 
Integration of nonlinear equations of mathematical physics by the method of inverse scattering, 
{\em Funct. Anal. Appl.} {\bf 13} (1979), 166--174.

\bibitem{Olv77}
P.J. Olver,
Evolution equations possessing infinitely many symmetries,
{\em J. Math. Phys.} {\bf 18} (1977), 1212--1215.

\bibitem{symmintegr1}
V.V. Sokolov and A.B. Shabat, 
Classification of integrable evolution equations,
{\em Sov. Sci. Rev. C} {\bf 4} (1984), 221--280.

\bibitem{symmintegr2}
A.V. Mikhailov, A.B. Shabat and R.I. Yamilov R I,
The symmetry approach to classification of integrable equations: complete lists of integrable systems,
{\em Russ. Math. Surv.} {\bf 42} (1987), 1--51.

\bibitem{symmintegr3}
A.S. Fokas, 
Symmetries and integrability, 
{\em Stud. Appl. Math.} {\bf 77} (1987) 253--299.

\bibitem{symmintegr4}
A.V. Mikhailov, A.B. Shabat and V.V. Sokolov, 
Symmetry approach to classification of integrable equations, 
in {\em What is Integrability?}, 
%Editor: Zakharov V E, 
Springer-Verlag, 1999. 

\bibitem{SanWan98}
J.A. Sanders and J.-P. Wang, 
On the integrability of homogeneous scalar evolution equations, 
{\em J. Differential Equations} {\bf 147(2)} (1998), 410--434. 

\bibitem{SanWan00}
J.A. Sanders and J.-P. Wang, 
On the integrability of non-polynomial scalar evolution equations, 
{\em J. Differential Equations} {\bf 166} (2000), 132--150.

\bibitem{Ver}
J.M. Verosky,
Negative powers of Olver recursion operators,
{\em J. Math. Phys.} {\bf 32} (1991), 1733--1736.

\bibitem{ZakManNovPit}
V.E. Zakharov, S.V. Manakov, S.P. Novikov, L.P, Pitaevsky, 
{\em Soliton Theory: The Method of Inverse Problem}, 
Nauka (Moscow) 1980. 

\bibitem{Fuc96}
B. Fuchssteiner, 
Some tricks from the symmetry-toolbox for nonlinear equations: generalizations of the Camassa-Holm equation,
{\em Physica D} {\bf 95} (1996), 229-–243.

\bibitem{AblKauNewSeg}
M.J. Ablowitz, D.J. Kaup, A.C. Newell, H. Segur,
The inverse scattering transform: Fourier analysis for nonlinear problems, 
{\em Stud. Appl. Math.} {\bf 53} (1974), 249--315.

\bibitem{CalDeg}
F. Calogero and A. Degasperis,
Nonlinear evolution equations solvable by the inverse spectral transform,
{\em Il Nuovo Cimento B} {\bf 32} (1976) 201--242. 

\bibitem{Qia}
Z. Qiao,
A general approach for getting the  commutator representations of the hierarchies of nonlinear evolution equations, 
{\em Physics Letters A} {\bf 195} (1994) 319--329.

\bibitem{QiaCaoStr}
Z. Qiao, C. Cao, W. Strampp,
Category of nonlinear evolution equations, algebraic structure, and r-matrix,
{\em J. Math. Phys.} {\bf 44} (2003), 701--722.

\bibitem{GomFraMelZim}
J.F. Gomes, G. Starvaggi Franca, G.R. de Melo, A.H. Zimerman,
Negative even grade mKdV hierarchy and its soliton solutions, 
{\em J. Phys. A:Math. Theor.} {\bf 42} (2009) 445204 (11pp).

\bibitem{FraGomZim}
G. Starvaggi Franca, J.F. Gomes, A.H. Zimerman,
The algebraic structure behind the derivative nonlinear Schrodinger equation, 
{\em J. Phys. A:Math. Theor.} {\bf 46} (2013) 305201 (19pp).

\bibitem{Anc03}
S.C. Anco, 
Conservation laws of scaling-invariant field equations, 
{\em J. Phys. A: Math. and Gen.} {\bf 36} (2003), 8623--8638.

\bibitem{Wan} 
J.-P. Wang, 
A list of 1+1 dimensional integrable equations and their properties, 
{\em J. Nonlinear Math. Phys.} {\bf 9} (2002), 213--233. 

\bibitem{QiaStr}
Z. Qiao, W. Strampp,
Negative order MKdV hierarchy and a new integrable Neumann-like system,
{\em Physica A} {\bf 313} (2002), 365--380.

%K, SK cons laws
\bibitem{FucOev}
B. Fuchssteiner and W. Oevel,
The bi-Hamiltonian structure of some nonlinear fifth- and seventh- order differential equations and recursion formulas for their symmetries and conserved covariants, 
{\em J. Math. Phys. } {\bf 23} (1982), 358--363. 

\bibitem{QiaLi}
Z. Qiao, J. Li,
Negative-order KdV equation with both solitons and kink wave solutions,
{\em Euro. Phys. Lett.} {\bf 94} (2011), 50003 (5pp).

%KK cons laws
\bibitem{SatKau}
J. Satsuma and D.J. Kaup,
A Backlund transformation for a higher order Korteweg-De Vries equation,
{\em J. Phys. Soc. Japan} {\bf 43} (1977) 692--697.

\bibitem{OlvRos}
P.J. Olver and P. Rosenau,
Tri-Hamiltonian duality between solitons and solitary-wvae solutions having compact support,
{\em Phys. Rev. E} {\bf 53} (1996) 1900--1906. 


\end{thebibliography}
\end{document}